\documentclass[prb,twocolumn,superscriptaddress,showpacs,amsmath,amssymb,longbibliography]{revtex4-2}
\usepackage[utf8]{inputenc}
\usepackage{graphicx}
\usepackage{bm}
\usepackage[usenames,dvipsnames]{xcolor}

\definecolor{goodred}{rgb}{0.7,0,0}
\usepackage[colorlinks,urlcolor=blue,citecolor=blue,linkcolor=goodred]{hyperref}

\graphicspath{{./figs/}}

\renewcommand{\vec}[1]{\bm{#1}}
\newcommand{\dd}[1]{d{#1}\,}

\DeclareMathOperator{\tr}{tr}
\newcommand{\diag}{\mathop{\mathrm{diag}}}

\newcommand{\expval}[1]{\left\langle{#1}\right\rangle}
\newcommand{\disp}[1]{{\varepsilon_{\vec{#1}}}}

\usepackage{verbatim}

\def\arxivversion{}

\begin{document}

\title{Superconducting junctions with flat bands}

\author{P. Virtanen}
\email{pauli.t.virtanen@jyu.fi}
\affiliation{Department of Physics and Nanoscience Center, University of Jyväskylä, P.O. Box 35 (YFL), FI-40014 University of Jyv\"askyl\"a, Finland}

\author{R. P. S. Penttil\"a}
\affiliation{Department of Applied Physics, Aalto University School of Science, FI-00076 Aalto, Finland}

\author{P. T\"orm\"a}
\affiliation{Department of Applied Physics, Aalto University School of Science, FI-00076 Aalto, Finland}

\author{A. D\'iez-Carl\'on}
\affiliation{Fakult\"at f\"ur Physik, Ludwig-Maximilians-Universit\"at, Schellingstrasse 4, 80799 M\"unchen, Germany}
\affiliation{Munich Center for Quantum Science and Technology (MCQST), M\"unchen, Germany}

\author{D. K. Efetov}
\affiliation{Fakult\"at f\"ur Physik, Ludwig-Maximilians-Universit\"at, Schellingstrasse 4, 80799 M\"unchen, Germany}
\affiliation{Munich Center for Quantum Science and Technology (MCQST), M\"unchen, Germany}

\author{T. T. Heikkil\"a}
\affiliation{Department of Physics and Nanoscience Center, University of Jyväskylä, P.O. Box 35 (YFL), FI-40014 University of Jyv\"askyl\"a, Finland}

\date{\today}

\begin{abstract}
  We analyze the properties of flat-band superconductor junctions that behave differently from ordinary junctions containing only metals with Fermi surfaces. In particular, we show how in the tunneling limit the critical Josephson current between flat-band superconductors is inversely proportional to the pair potential, how the quantum geometric contribution to the supercurrent contributes even in the normal state of a flat-band weak link, and how Andreev reflection is strongly affected by the presence of bound states. Our results are relevant for analyzing the superconducting properties of junctions involving electronic systems with flat bands.
\end{abstract}

\maketitle

\begin{figure}[t]
  \includegraphics{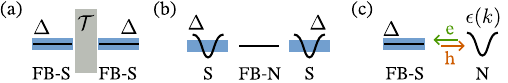}
  \caption{
    \label{fig:sche}
    Junctions between flat-band and dispersive superconductors (with order parameter $\Delta$)
    and normal-state systems.
    (a)~Superconducting flat-band (FB-S) tunnel junction.
    (b)~Normal flat-band (FB-N) system between superconductors (S).
    (c)~Andreev reflection in flat-band superconductor / normal metal (N) junction.
  }
\end{figure}

Most of our knowledge of the properties of superconducting junctions relies on an expansion of the electronic dispersion around the Fermi energy of the electrodes~\cite{bardeen1957,andreev1964,josephson1962,giaever1961study,ambegaokar1982-qdo,beenakker1991universal,beenakker1992quantum}. This expansion is often coined {\it Andreev approximation} to indicate its use in describing the Andreev reflection that is at the heart of generating essentially all relevant effects encountered on those junctions: the proximity effect and the supercurrent, the excess current and the multiple Andreev reflections. Since the Andreev approximation relies on the presence of the Fermi surface, it completely fails for systems that lack it, such as systems containing flat electronic bands. Flat bands are promising for increasing the critical temperature~\cite{kopnin2011} and providing unconventional superconductivity~\cite{andrei2021} with quantum geometric effects~\cite{torma2022}, therefore the behavior of superconducting junctions in the flat band limit has become an important open question. Systematic investigation of materials \cite{jiang2024twistable} may also bring new flat-band superconductors to light.

In this Letter, we study how the flat-band nature of the bands modifies the superconducting transport properties of superconducting junctions. The most relevant changes are due to the localization of the energy window relevant to the transport properties and the possibility of supercurrent even in the presence of frozen quasiparticle transport~\cite{pyykkonen2023-snq}. The latter arises in the case of non-trivial quantum geometry of the flat-band eigenfunctions~\cite{peotta2015}. Moreover, as the electronic states in flat band systems are inherently localized, many transport properties can be linked to the coupling between bound states. In contrast, in ordinary junctions such bound-state corrections can often be neglected. We illustrate these effects with three main results: (i) The critical current of a tunnel junction between flat-band superconductors is inversely proportional to the pair potential, in contrast to the linear dependence in ordinary junctions. (ii) Supercurrent through a flat-band weak link depends crucially on the strength of interactions inside that weak link. Finally, (iii) Andreev reflection from a flat-band superconductor depends on surface bound-state contributions that also affect the excess current across the junction. The limit of an exact flat band is a useful idealization. Using example models, we discuss how this limit emerges for the above three properties as a dispersive band becomes flatter. We also identify the relevant limiting scales describing the crossover between ordinary and flat bands. 

\paragraph*{Tunnel junctions.}

\begin{figure}
  \includegraphics{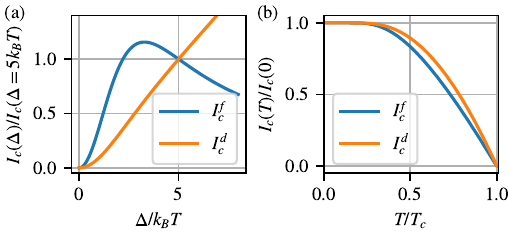}
  \caption{
    \label{fig:AB}
   (a) The flat-band and dispersive Josephson currents as a function of $\Delta/(k_B T)$. (b) The flat-band and dispersive Josephson currents as a function of temperature. The superconducting gap $\Delta(T)$ is solved self-consistently with BCS theory. 
    }
\end{figure}

First, we show that the presence of flat bands can already be identified from the parameter dependence of the critical
current in tunnel junctions [see Fig.~\ref{fig:sche}(a)].  The Ambegaokar--Baratoff formula for
the Josephson current through two superconductors connected with a
simple tunnel junction is the simplest description of such junctions. For two single-band superconductors with equal amplitude
superconducting gaps, the zero-voltage Josephson current is given by
\cite{ambegaokar1982-qdo, mahan00}
\begin{align}
  \label{eq:AB}
    I_s=\sin(\phi)I_c=\sin(\phi)\frac{4e|\mathcal{T}|^2}{\hbar} k_B T\sum_{i\omega}\sum_{\vec{k}\vec{p}}F^\dagger_{\vec{k}, i\omega}F_{\vec{p}, i\omega'},
\end{align}
where $\phi$ is the phase difference between the superconducting gaps, $I_c$ is the critical current, $\mathcal{T}$ is the tunneling amplitude, $e$ is the charge of electrons, $T$ is the temperature, $k_B$ the Boltzmann constant, and $F_{\vec{k},i\omega} = \frac{\Delta}{\varepsilon_{\vec{k}}^2+\Delta^2 + \omega^2}$ denotes the anomalous Bardeen-Cooper-Schrieffer (BCS) Green's function. Here, $\omega$ is the Matsubara frequency, and $\Delta$ is the absolute value of the superconducting gap. The momentum summed Green's functions are different for the regular quadratic dispersion $\varepsilon_{\vec{k}}^d = \hbar^2\vec{k}^2/2m-\mu$, where $\mu$ is the chemical potential, and a flat dispersion $\varepsilon_{\vec{k}}^f=0$:
\begin{align}
    \sum_{\vec{k}} F^d_{\vec{k},i\omega} = \nu_F\pi \frac{\Delta}{\sqrt{\omega^2+\Delta^2}}, \;\;\;\;\;  \sum_{\vec{k}} F^f_{\vec{k},i\omega} = \frac{\Delta}{\omega^2+\Delta^2},
\end{align}
where $\nu_F$ is the normal-state Fermi-level density of states of the dispersive leads (with units 1/energy). Here, we consider a completely flat band in the whole Brillouin zone. The critical Josephson currents are given by
\begin{align}
    I_c^d &= I_{c0}^d \tanh \tilde \Delta,\quad I_{c0}^d = 2e |\mathcal{T}|^2 \pi^2 \nu_F^2 \Delta/\hbar \label{eq:ABn}\\
    I_c^{fb} &= I_{c0}^{fb} \left[\tanh \tilde \Delta - \tilde \Delta {\rm sech}^2\tilde \Delta\right], \quad I_{c0}^{fb} = \frac{e|\mathcal{T}|^2 }{\hbar\Delta} \label{eq:ABfb}
\end{align}
for the dispersive and flat-band cases, respectively. Here $\tilde \Delta = \Delta/(2 k_B T)$. 
A notable qualitative difference is the dependence of the zero-temperature critical current on $\Delta$: in the ordinary case $I_c$ is linearly proportional to $\Delta$, whereas, for a flat band, it is inversely proportional to it, up to $\Delta\sim{}k_BT$ as shown in Fig.~\ref{fig:AB}. In contrast, the current between a dispersive and a flat-band lead has no dependence on $\Delta$ in the zero-temperature limit \cite{supplement}. The tunneling over the insulating barrier is mediated by virtual pair breaking with probability $|\mathcal{T}|^2/\Delta$, which explains the $I_{c0}^{fb} \propto 1/\Delta$ behavior in a flat band where all electrons participate in pairing at any value of $\Delta$. In the dispersive case, the fraction of pairs increases with increasing $\Delta$ on both sides of the junction, leading to $I_{c0}^d \propto \Delta^2/\Delta = \Delta$. In the case of weakly dispersive bands with bandwidth $J$, the relative correction to Eq.~\eqref{eq:ABfb} is proportional to $J^2/\Delta^2$ \cite{supplement}, and hence we expect Eq.~\eqref{eq:ABfb} to hold for $J \ll \Delta$. To clearly distinguish the dependencies on $\Delta$, its value close to the junction should be varied, for example via magnetic field or supercurrent flow close to the tunnel interface \cite{anthore2003,ligato2022-tsq}, while keeping $T$ fixed.

Another way to give an intuitive understanding of the FB tunneling
current is to consider the coupling of localized states $a$ and $b$,
representing the flat bands, on opposite sides of the tunnel
junction. The simplest description in the spin-degenerate case is the
Hamiltonian $H= t a_\uparrow^\dagger b_\uparrow + t
a_\downarrow^\dagger b_\downarrow + \Delta a_{\uparrow}^\dagger
a_{\downarrow}^\dagger + \Delta e^{i\varphi} b_{\uparrow}^\dagger
b_{\downarrow}^\dagger + \mathrm{h.c.}$ where $t$ is the tunnelling
amplitude.  It has two Andreev bound states,
$\epsilon_\pm=\sqrt{|\Delta|^2+|t|^2 \pm 2|t \Delta|
  \sin(\varphi/2)}$ \footnote{Note that the bound state energies
differ from the usual point contact result obtained in
\cite{beenakker1991universal}.}.  The supercurrent flowing between the
two superconductors is then $I_S(\varphi)=-\frac{2e}{\hbar}N\sum_\pm
\frac{\partial\epsilon_\pm}{\partial\varphi}\tanh\frac{\epsilon_\pm}{2k_BT}$,
where $N$ is the total number of states participating in
tunneling~\cite{supplement}.
Note that the result differs from the dispersive superconductor point-contact current \cite{beenakker1991universal}.
For $t\to0$, this reduces to
Eq.~\eqref{eq:ABfb} with $N|t|^2=|\mathcal{T}|^2/(2\pi)$. The
$1/\Delta$ divergence of the supercurrent at $\Delta\sim{}k_BT\to0$ is
cut off at $\Delta\simeq{}|t|$, where the critical current saturates
to $I_c^{fb}\sim\frac{e\Delta}{\hbar}N$.

\paragraph*{Lattice models.}

\begin{figure}
  \includegraphics[width=\columnwidth]{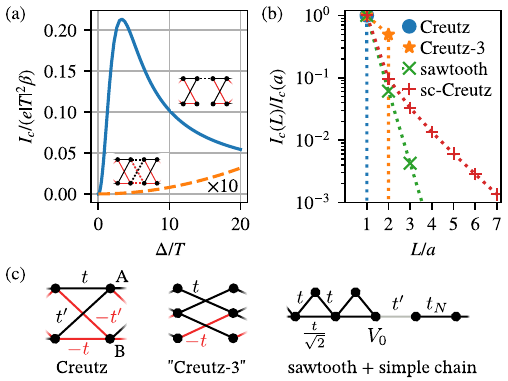}
  \caption{
    \label{fig:latt}
    (a)
    Dependence of the supercurrent in a flat-band tunnel junction on the interface coupling
    between two semi-infinite Creutz ladders.
    Solid: tunneling only between A-sites.
    Dashed: tunnel hoppings $\propto$ bulk hoppings. In both cases, tunnel hoppings $|T|\ll{}|t|$.
    Here $\beta$ is the inverse temperature.
    (b)
    S/N/S junction scaling of the supercurrent vs.~length of N, for different
    lattice types.
    Here ``sc-Creutz'' is the Creutz ladder with attractive interaction in the N-region, see Fig.~\ref{fig:sche}(b).
    (c)
    1D lattices with flat bands, with hoppings indicated.
    For the sawtooth ladder, the connection to the simple chain with an edge potential
    $V_0$ is illustrated.
  }
\end{figure}

To extend the AB results beyond simple tunneling, we consider
tight-binding models for the junction, with nontrivial sublattice
structure and taking edge effects into account.  In this case,
Eq.~\eqref{eq:AB} generalizes to~\cite{martinrodero1994,supplement}
\begin{align}
  \label{eq:Icgeneral}
  I_S = i T \sum_{\omega_n} \tr(G_{LR} J \tau_3 - G_{RL} J^\dagger\tau_3)
  \,,
\end{align}
where $J=J_0e^{-i\varphi\tau_3/2}$ is the tunnel hopping matrix including
the phase difference, $G_{LR/RL}$ are elements of the full Green's
functions connecting the two sides, and $\tau_3$ is a Pauli matrix in the Nambu space. 

In point contacts between flat-band materials with sublattice
structure, one can express the current as a sum of several
contributions of the same type as in the two-site tunneling model, but
with effective tunnel amplitudes determined by the contact details
\cite{supplement}.
The supercurrent is then sensitive to
the structure of the tunnel interface. An extreme fine-tuned case is tunnel
hoppings proportional to the bulk hopping matrix $J$ between unit
cells in 1D chains, which results in the cancellation
of the leading contribution to the supercurrent found in the AB
result \cite{supplement}. This is illustrated in Fig.~\ref{fig:latt}(a) for two
semi-infinite Creutz ladders (Fig.~\ref{fig:latt}(c)) coupled at the
ends: The flat-band peak of Eq.~\eqref{eq:ABfb} is visible in the case
where the tunnel coupling between unit cells at the ends is only
between A-sites (solid line), whereas in the case of coupling that follows the hopping structure of the ladder but with a smaller magnitude ($\alpha{}J$ with $\alpha<1$), the critical current resembles the
dispersive band result (dashed). \cite{codes}

\paragraph*{Superconducting--normal junctions.}

In junctions where the tunneling weak link is replaced by an extended
normal-state (N) layer, generally, the proximity effect and the
magnitude of the supercurrent decays as the junction length
increases~\cite{likharev1979superconducting,golubov2004} on a
coherence length scale determined by the Fermi velocity.  When the
N-layer is a flat-band system with low Fermi velocity, a natural
question is whether the decay length scale would be given by the
quantum geometry, analogously to the superfluid weight in the
superconducting state \cite{peotta2015,torma2022}. To address this, we
first make a simple argument showing that in general no such analogy
holds for long junctions, and the decay length depends on other
model-dependent factors.  However, at short distances, there is a
connection to a coherence length defined as the root mean square
of the pair correlation function. Finally, we show how weak attractive interactions can further
modify the conclusions.

First, consider a simple argument to estimate the spatial behavior,
in 1D and quasi-1D tight-binding models with finite range hopping, along the lines of Ref.~\cite{chen2014-ief}.
The spatial extent of the proximity effect and supercurrent decay can be characterized by
the pair correlation function $\Pi(x,x')$~\cite{larkin2005},
which describes the propagation
of a pair of electrons from $x$ to $x'$. 
Here, $x=(m,\alpha)$ specifies both unit cell position $m$ and orbital
$\alpha$ indices.  For a weak proximity effect, it can be written as
$\Pi_{\alpha\alpha'}(m,m')=T\sum_{\omega_n}G_{\alpha\alpha'}(m-m',\omega_n)G_{\alpha\alpha'}(m-m',-\omega_n)$,
where $G$ is a normal Green's function.  In the finite range hopping (quasi-)1D models, the
Bloch Hamiltonian $H(k)=\sum_{j=-M}^M H_j e^{ijka}$ is a Laurent
polynomial in $e^{ika}$ of order $M$ (range of hopping). Here $a$ is
the lattice constant.  From Cramer's rule, it then follows that the
bulk Green's function $G(k,\omega) = [i\omega-H(k)]^{-1}$ can be
written as
\begin{align}
  \label{eq:Gk}
  G(k,\omega) = \frac{\sum_{j=-(N-1)M}^{(N-1)M}e^{ijka}P_{\omega,j}}{\prod_p[i\omega - \epsilon_p(k)]}
  \,,
\end{align}
where $\epsilon_p(k)$ are the band dispersions of the N-material for bands $p$, and
the numerator is a matrix Laurent polynomial of order $\le(N-1)M$,
which depends on the number $N$ of orbitals per unit cell. Taking the inverse Fourier
transform, in real space we find
\begin{align}
  \label{eq:Gm}
  G(m,\omega)
  &=
  \sum_{j=-(N-1)M}^{(N-1)M} P_{\omega,j} f_\omega(m-j)
  \,,
  \\
  f_{\omega}(j)
  &=
  \int_{-\pi/a}^{\pi/a}\frac{dk \, ae^{-ijka}}{2\pi\prod_p[i\omega - \epsilon_p(k)]}
  \,.
\end{align}
If all bands are exactly flat, then $f(j)\propto\delta_{j0}$, and
$G(m,\omega)=0$ for $|m|>(N-1)M$.
Thus, in models with only exact flat
bands, compact localized states may carry current up to the distance
$\xi_{\rm max}=(N-1)Ma$, but at longer distances the correlations
vanish. This applies also to the Bogoliubov - de Gennes (BdG) Hamiltonian, as observed in Ref.~\onlinecite{thumin2024-cfc}
in specific models.
More generally, at distances larger than $\xi_{\rm
  max}$ the decay is described by $f$, which depends \emph{only} on
the band dispersions $\epsilon_p(k)$~\footnote{In quasi-1D, dispersions of possible edge states also contribute here. }.
With a
dispersive band at the Fermi level, this produces the exponential decay on
the scale of the ballistic coherence length $\xi_{v_F}\propto{}\hbar{}v_F/(2\pi
T)$ defined by the Fermi velocity~\cite{golubov2004}.
On the other hand, for a flat band,
the long-distance behavior is similar to evanescent decay in insulators:
temperature-independent and governed by the energy gap to the nearest dispersive band; we denote this coherence length by $\xi_{\rm gap}$.

The short-distance behavior, however, depends not only on band dispersions
but also on the band geometry.  Define the root mean square (rms) range
of the proximity effect as
\begin{align}
  (\delta x)_{\alpha\alpha',\mathrm{rms}}^2
  \equiv
  \frac{\sum_m \Pi_{\alpha\alpha'}(m,0) a^2m^2}{2\sum_m \Pi_{\alpha\alpha'}(m,0)}
  =
  \frac{-\partial_q^2\Pi_{\alpha\alpha'}(q)\rvert_{q=0}}{2\Pi_{\alpha\alpha'}(q)\rvert_{q=0}}
  \,.
\end{align}
Under assumptions of time reversal symmetry and uniform pairing on an
isolated flat band, the right-hand side can be related to the
Brillouin-zone average of the minimal quantum metric $g(k)$, via $(\delta
x)_{\alpha\alpha',\mathrm{rms}}^2 = \int\frac{a\,dk}{2\pi} g(k)$~\cite{chen2024,hu2024anomalouscoherencelengthsuperconductors,iskin2024-cld}. 
A similar relation can also be found for
the Cooper pair rms size~\cite{iskin2024pairsizequantumgeometry}.
However, as the decay of the long-distance tail of $\Pi$ is governed by the band dispersions, a general relation between the critical current and quantum geometry as
suggested in Ref.~\onlinecite{li2024-fbj} probably does not exist.

Figure~\ref{fig:latt}(b) illustrates the above results, showing the dependence of the critical current
$I_c$ of an S/N/S junction as a function of the N-region length. Here, N
is constructed from the same lattice as S [shown in
  Fig.~\ref{fig:latt}(c)], but with $\Delta=0$ on $n=L/a-1$ lattice
sites.  The Creutz (taking $t=t'$) and the ``Creutz-3'' lattices have
exact flat bands, and supercurrent vanishes exactly after a finite
number of sites.  In contrast, the sawtooth lattice has both flat and
dispersive bands, which allows for the exponential decay in the
critical current as a function of $L$ at a scale determined by the band dispersion.

\begin{figure}
  \includegraphics[width=\columnwidth]{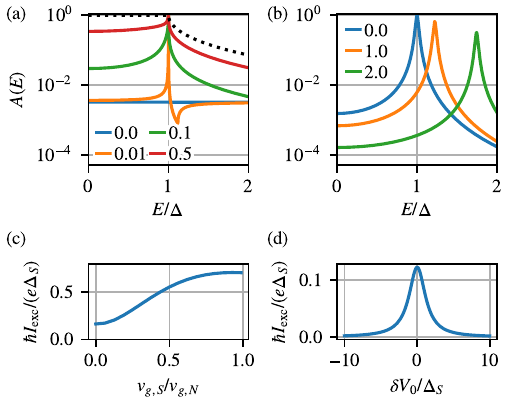}
  \caption{
    \label{fig:AR}
    (a) Andreev reflection probability in a FB-S/N junction from Eq.~\eqref{eq:scattering},
    for different ratios of $v_{g,S}/v_{g,N}$.
    The dotted line corresponds to the dispersive BTK result for a transparent interface.
    (b)
    Resonant Andreev reflection from a sawtooth lattice edge state,
    for different edge potential offsets $\delta V_0$, and tunnel coupling $t'=t/10$.
    (c,d) Corresponding excess current.
  }
\end{figure}

\paragraph{Interaction-mediated supercurrent.}

Vanishing of supercurrent due to localization of quasiparticles in the
N region may be lifted by a small attractive interaction.  With
nontrivial quantum geometry, the superfluid weight is nonzero even if
bands are exactly flat, allowing for Cooper pair transport~\cite{peotta2015,torma2022}.  Importantly,
we find that in an S/N/S geometry, the interaction strength
in N does not need to be large enough to make N an intrinsic superconductor. This component of
the SNS supercurrent is often neglected in systems with dispersive
bands, but in localized flat band systems it is the only long-distance
contribution.

Starting with a simple example, in Fig.~\ref{fig:latt}(b) we show a result for the Creutz ladder
with small attractive interaction. This is from a mean-field
model with local superconducting pair potential
$\Delta_i=U\langle{c_{i\uparrow}c_{i\downarrow}}\rangle$ on each site of the N-region,
taking $\mu=-2t$ centered at the lower flat band.
For any $U<U_c(T)$ below the interaction required for intrinsic superconductivity,
the supercurrent decays exponentially with the system length, $I_c\propto{}e^{-L/L_g}$ ($L_g$ is defined below).
For $U=0$, there is no supercurrent for $L>a$.

More generally, the Ginzburg--Landau (GL) expansion \cite{larkin2005,chen2024} captures the behavior of the interaction-induced component of
SNS critical current.  Given (quasi-)1D GL free energy
density for the N-region,
\begin{align}
  \label{eq:F-gradexp}
  F = \frac{1}{2}\mathcal{A} D(T)|\partial_x\Delta|^2 + \mathcal{A} b(T)|\Delta|^2
  \,,
\end{align}
where $\mathcal{A}$ is the cross-sectional area (3D) or width (2D) of the junction,
the interaction-mediated critical current is, in long junctions $L>L_g$,
\begin{align}
  I_{c,\rm int} \simeq 8 \mathcal{A} b(T) \Delta_S^2 L_g e^{-L/L_g}
  \,,
\end{align}
where $L_g=\sqrt{D(T)/[2b(T)]}$, and $\Delta_S$ is the pair potential of the S-leads.
The GL expansion assumes $\Delta_S\ll{}T$. Importantly, the result can also be nonzero if the bands are exactly flat.
For the above Creutz ladder model at $\mu=-2t$, $t=t'$ (and assuming the orbitals A and B coincide spatially which makes the quantum metric equal the minimal quantum metric~\cite{huhtinen2022,Tam2023}), the expansion gives
$D=\frac{\xi_g^2}{4T}$, $b=\frac{2}{U} - \frac{1}{4T}$, $T_c=U/8$.
The superfluid weight $D$ and the decay length scale $L_g$ are proportional
to the average of the (minimal) quantum metric tensor $g^{(0)}$ of the flat band,
\begin{align}
  L_g &= \xi_g \sqrt{\frac{T_c}{T-T_c}}
  \,,
  &
  \xi_g^2 &= a \int_{-\pi/a}^{\pi/a} \frac{dk}{2\pi} g^{(0)}(k)
  \,.
\end{align}
The Creutz ladder GL problem can also be solved without the gradient expansion in Eq.~\eqref{eq:F-gradexp}~\cite{supplement}, which results in $I_{c,\rm int}=\frac{\Delta_S^2}{2T}\sqrt{z(z-1)}e^{-L/L_g}$,
$L_g=a/\log[2z-1+2\sqrt{z(z-1)}]$ where $z=T/T_c>1$. 

Notably, $I_{c,\rm int}$ does not depend on the normal state resistance unlike the conventional SNS critical current~\cite{likharev1979superconducting,dubos2001josephson}. Such a flat-band contribution could explain the anomalous relationship between the critical current and normal state resistance observed in recent experiments on magic-angle twisted bilayer graphene Josephson junctions~\cite{Diez2025}. 

The different decay length scales could be
distinguished in experiments by their different temperature dependence. In
particular, the Fermi surface coherence length scale
$\xi_{v_F}\propto{}\hbar{}v_F/(2\pi T)$ is strongly temperature-dependent,
as is the interaction-driven $L_g$, whereas the root mean square spread of the pair correlation function
$(\delta{}x)_{\alpha\alpha',\mathrm{rms}}$ and $\xi_{\rm gap}$ given by dispersive bands separated by a gap from the band of interest do not
vary strongly with temperature.

\paragraph*{Andreev reflection.}
In the textbook Andreev reflection calculation~\cite{blonder1982-tfm},
one matches incoming waves at energy $\epsilon$ to evanescent states
inside the superconductor at the same energy.  For the exact flat band
limit, the dispersion of the propagating modes (real $\vec{k}$) is a
constant, i.e.~$\epsilon(\vec{k})=\epsilon_n$, and analytic
continuation implies absence of evanescent modes for any complex
$\vec{k}$ with $|\vec{k}|<\infty$ at energies
$\epsilon\ne\epsilon_n$. In lattice models with exact flat bands, the
evanescent modes vanish exactly after a finite number of
lattice sites \cite{supplement}.

The (quasi-)1D tight-binding Andreev scattering problem can be directly solved.
For an S/N interface where N consists of 1D chains with hopping
$t_N$ at half filling connected to each edge site of S, the backscattering matrix is \cite{supplement}
\begin{align}
  R = - \frac{1 - g t_N e^{ik_e}}{1 - g t_N e^{-ik_e}}
  \,,
  \label{eq:scattering}
\end{align}
where $k_e=\arccos(\epsilon/2t_N)$, and $g$ is the Bogoliubov--de Gennes (BdG) surface Green's function of S.
The matrix relates the incoming and outgoing scattering modes in the N lead to each other,
$a_{R,\rm out}=R a_{R,\rm in}$, and Andreev reflection amplitudes are found in its Nambu off-diagonal components. %

The Fermi velocity mismatch between flat-band S and N acts similarly
to the barrier height in the Blonder-Tinkham-Klapwijk (BTK) model
\cite{blonder1982-tfm}.  The Andreev reflection probability for the
Creutz ladder as a FB-S [Fig.~\ref{fig:sche}(c)] is illustrated in
Fig.~\ref{fig:AR}(a), for $t=t_N=50\Delta\gg\Delta$, $\mu$ tuned at
the flat band, and different values of $t'$ allowing to tune the ratio
of group velocities $v_{g,S}/v_{g,N}$ on the two sides of the
interface. This illustrates the cross-over from a dispersive to an
exact flat band on the S-side. When approaching the exact flat-band
limit, the resonance at $E=\Delta$ gradually disappears and converges
to a slowly varying background determined by the matching of the
propagating modes of N to the localized evanescent states of S in its
normal state. This behavior is similar for different lattice models
with a mismatch at the interface.

Flat band systems may additionally host edge states, visible
as additional poles in $g$ in Eq.~\eqref{eq:scattering}. If the edge
state energies are close to the chemical potential, they provide a
resonant Andreev reflection peak in the reflection probability.  These
resonances generally can be visible in transport properties
\cite{pyykkonen2021,pyykkonen2023-snq}. To illustrate,
in Fig.~\ref{fig:AR}(b) we show the Andreev reflection
probability for the sawtooth lattice [cf.~Fig.~\ref{fig:latt}(c)],
tuning the edge potential $V_0=-t/\sqrt{2} + \delta V_0$
to align the edge state with the chemical potential $\mu=-t\sqrt{2}$.

\paragraph*{Excess current.}
The energy dependence of the Andreev reflection probability is accessible via $dI/dV$ measurements of N/S junctions, and in the excess current
\cite{blonder1982-tfm}, $I_{\rm exc} =
e\int_{-\infty}^{\infty}\dd{\epsilon}\nu v_{g} [A(\epsilon,\Delta) -
  B(\epsilon,\Delta) + B(\epsilon,0)]$, where $A$ and $B$ are the Andreev and normal
reflection probabilities and $\nu$ and $v_{g}$ the density of states and
group velocity on the normal side. These are shown in Fig.~4(c,d)
corresponding to the cases considered in Fig. 4(a,b). They
illustrate the general expectation that Andreev reflection in the strict flat-band limit differs
significantly from dispersive superconductors, and is sensitive to the interface matching.

\paragraph*{Conclusions.} We have shown that the characteristics of Josephson junctions are drastically affected by flat bands. The tunneling junction between two flat-band superconductors shows a critical current inversely proportional to the order parameter $\Delta$, opposite to the linear in $\Delta$ dependence in the conventional dispersive case. In S/N/S junctions, we observe that interactions in the N-region
become increasingly important as the dispersion in N flattens. When N
has exactly flat bands, supercurrent is only carried by the
interaction component at long distances. Finally, we found that the Andreev reflection and excess current in flat band systems are highly sensitive to the properties of the edge states at the interfaces. These results will be useful in interpreting flat-band experiments, as already indicated in the case of twisted bilayer graphene in Ref.~\cite{Diez2025}, and inspire superconducting devices with new functionalities.

\begin{acknowledgments}
\paragraph*{Acknowledgments} R.P.~acknowledges financial support from the Fortum and Neste Foundation. D.K.E. acknowledges funding from the European Research Council (ERC) under the European Union’s Horizon 2020 research and innovation program (grant agreement No. 852927), the German Research Foundation (DFG) under the priority program SPP2244 (project No. 535146365), the EU EIC Pathfinder Grant “FLATS” (grant agreement No. 101099139) and the Keele Foundation as part of the SuperC collaboration. This work was supported by the Research Council of Finland under project numbers 339313 and 354735, by European Union’s HORIZON-RIA programme 331 (Grant Agreement No.~101135240 JOGATE), and by Jane and Aatos Erkko Foundation, Keele Foundation, and Magnus Ehrnrooth Foundation as part of the SuperC collaboration.
\end{acknowledgments}

\bibliography{refs,references}

\ifx\arxivversion\undefined
\else
\clearpage
\appendix
\setcounter{figure}{0}
\setcounter{equation}{0}
\renewcommand\thefigure{S\arabic{figure}}
\renewcommand\theequation{S\arabic{equation}}

\graphicspath{{./figs/}}

\section{Ambegaokar--Baratoff formula for flat bands}

 We begin with the tunneling Hamiltonian
\begin{equation}
    H = H_R + H_L + H_T =H_R + H_L + \sum_{\vec{k}\vec{p}\sigma}\mathcal{T}_{\vec{k}\vec{p}}c^\dagger_{\vec{k}\sigma}c_{\vec{p}\sigma} + \text{h.c.},
\end{equation}
where $H_R$ and $H_L$ describe the right and the left leads and $\mathcal{T}_{\vec{k}\vec{p}}$ is the tunneling matrix amplitude from the left lead at momentum $\vec{k}$ to the right lead at momentum $\vec{p}$. The tunneling current is given by $I(t) = -e \expval{\Dot{N}_L(t)}$, where $\Dot{N}_L$ is the change in the particle number of the left lead, which is obtained from the commutator between the Hamiltonian and the number operator of the left side, i.e., $\Dot{N}_L = i\left[H, N_L\right]$. The tunneling current comprises the single-particle current and the Josephson current. In the following, we focus on the Josephson current, which is given by \cite{mahan00}
\begin{equation}
    I_s(t) = \frac{2e}{\hbar} \text{Im}\left[e^{-2ietV/\hbar}\Phi(eV) \right],
\end{equation}
where $\Phi(eV)$ is the retarded correlation function 
\begin{equation}
   \Phi(eV) = -\int_{-\infty}^\infty dt\Theta(t-t')e^{ieV(t-t')}\left< \left[ A(t), A(t')\right]\right>.
\end{equation} Here, $A(t) = \sum_{\vec{k }\vec{p} \sigma}\mathcal{T}_{\vec{k}\vec{p}}c^\dagger_{\vec{k} \sigma}(t)c_{\vec{p} \sigma}(t)$ and $V=(\mu_l-\mu_r)/e$ is the electrical potential difference. The retarded correlation function can be evaluated by defining the Matsubara function $\Phi(i \omega)$ and later making the analytic continuation $i\omega \rightarrow eV+ i\delta$~\cite{mahan00}.
\begin{align}
    &\Phi(i \omega) = -\int_0^\beta d\tau e^{i\omega_n \tau} \expval{T_\tau A(\tau)A(0)}\\ 
    &= -2|\mathcal{T}|^2e^{-i\phi}\sum_{\vec{kp}}\underbrace{k_BT\sum_{i\nu}F^{l\dagger}_{\vec{k}, i\nu}F^r_{\vec{p}, i\nu-i\omega}}_{L_{\vec{kp}}(i\omega)} ,
\end{align}
where $F^{l/r}_{\vec{k}, i\omega}=-\frac{|\Delta|}{(i\omega)^2-\varepsilon_{\vec{k}}^2-|\Delta|^2}$ are the anomalous Green's functions of the left and right leads, expressed with the absolute values of the superconducting gaps $|\Delta|$, which are assumed to be equal for the leads. Here $\varepsilon_{\vec{k}}$ denotes the dispersion of the band and $\omega$ the Matsubara frequency. The phase difference between these superconducting gaps has been factored out of the summation, resulting in the term $e^{-i\phi}$. We also suppose that the tunneling matrix elements are independent of the momenta, i.e. $\mathcal{T}_{\vec{kp}}\mathcal{T}_{\vec{-k-p}} = |\mathcal{T}|^2e^{-i\phi_0}$, where $\phi_0=0$ in the presence of time-reversal symmetry in the junction. 

We use $L_{\vec{k}\vec{p}}(i\omega)$ to denote the result of the Matsubara summation of the anomalous Green's functions. Usually, the Matsubara summation should be made before the analytic continuation. However, we consider zero-voltage $V=0$, where we can do the analytic continuation $i\omega \rightarrow eV+ i\delta$ first and obtain

\begin{align}
    I_s(t) & = \frac{4e}{\hbar}\; \text{Im}\left[-e^{-i\phi}|\mathcal{T}|^2\sum_{\vec{kp}} L_{\vec{k}\vec{p}}(eV=0) \right] \\
    &= \text{sin}(\phi) \frac{4e}{\hbar}|\mathcal{T}|^2 k_BT\sum_{i\nu}\sum_{\vec{kp}}
    F^{l\dagger}_{\vec{k}, i\nu}F^r_{\vec{p}, i\nu}\\
    &= \text{sin}(\phi) I_c,
\end{align}
where the current amplitude is given by the critical current $I_c$.
The momentum summed Green's functions are different for the regular quadratic dispersion $\varepsilon_{\vec{k}}^d = \hbar^2\vec{k}^2/2m-\mu$ and a flat dispersion $\varepsilon_{\vec{k}}^f=0$:
\begin{align}
    \sum_{\vec{k}} F^d_{\vec{k},i\omega} = \nu_F\pi \frac{|\Delta|}{\sqrt{\omega^2+|\Delta|^2}}, \;\;\;\;\;  \sum_{\vec{k}} F^f_{\vec{k},i\omega} = \frac{|\Delta|}{\omega^2+|\Delta|^2},
\end{align}
where $\nu_F$ is the normal-state Fermi-level density of states of the dispersive leads. The critical Josephson currents for the flat-flat and dispersive-dispersive cases are obtained by performing the remaining Matsubara summation
\begin{align}
    I_c^d &= I_{c0}^d \tanh \tilde \Delta,\quad I_{c0}^d = 2e |\mathcal{T}|^2 \pi^2 \nu_F^2 |\Delta|/\hbar \label{eq:SABn}\\
    I_c^{fb} &= I_{c0}^{fb} \left[\tanh \tilde \Delta - \tilde \Delta {\rm sech}^2\tilde \Delta\right], \quad I_{c0}^{fb} = \frac{e|\mathcal{T}|^2}{\hbar|\Delta|} \label{eq:suppABfb}
\end{align}
for the dispersive and flat-band cases, respectively. Here $\tilde \Delta = |\Delta|/(2 k_B T)$. The current between a dispersive and a flat-band lead has to be calculated numerically from 
\begin{equation}
    I^{d-f}_c = \frac{4e |\mathcal{T}|^2 \nu_F\pi}{\hbar}  k_BT \sum_{i\omega} \frac{|\Delta|^2}{(\omega^2+|\Delta|^2)\sqrt{\omega^2+|\Delta|^2}},
\end{equation}
since the Matsubara summation does not have an analytical solution. Figure~\ref{fig:flat-disp current} shows the critical currents as a function of $|\Delta|$ in all three cases. For $\Delta \gg k_BT$, $I^{d}_c$ is linearly proportional to $\Delta$, and $I^{f}_c$ inversely, while $I^{d-f}_c$ seems to have almost no dependence on $\Delta$.
\begin{figure}
 \includegraphics{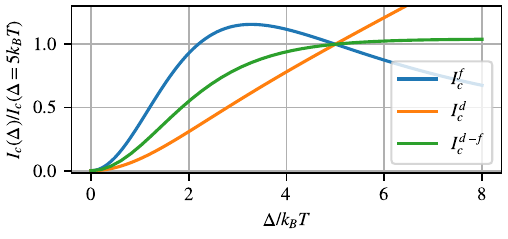}
  \caption{\label{fig:flat-disp current}
    The Josephson currents as a function of $\Delta/(k_B T)$ for the three possible junctions: dispersive-dispersive, flat-flat, and dispersive-flat.  
  }
  \end{figure}

We can examine the limit of the exact flat band approximation by Taylor expanding the anomalous Green's functions of both leads around zero dispersion $\disp{k} = 0$ and taking terms up to second order in $\disp{k}$
\begin{equation}
  F^{l/r}_{\vec{k}, i\omega}=\frac{|\Delta|}{\omega^2+\disp{k}^2+|\Delta|^2} \simeq \frac{|\Delta|}{\omega^2+|\Delta|^2}\left(1+\frac{\disp{k}^2}{\omega^2+|\Delta|^2}\right).
\end{equation}
Now, the critical current up to second order in $\disp{k}$ is given by
\begin{equation}
    \frac{I_c \hbar}{4e|\mathcal{T}|^2} \simeq k_BT\sum_{i\omega} \frac{|\Delta|^2}{(\omega^2+|\Delta|^2)^2} + 2\frac{E^2|\Delta|^2}{(\omega^2+|\Delta|^2)^3},
\end{equation}
where $E^2=\sum_{\vec{k}}\disp{k}^2$. The first term produces the flat-band current presented earlier and the second term gives the following second-order correction to the current
\begin{align}
   I^{(2)}_c &=  I_{c0}\left[\tanh\tilde{\Delta}-\tilde{\Delta}{\rm sech}^2\tilde{\Delta} - \frac{2\tilde\Delta^2}{3}{\rm sech}^2\tilde{\Delta}\tanh{\tilde{\Delta}}\right], \\
  I_{c0} &= \frac{3e|\mathcal{T}|^2 E^2}{2\hbar|\Delta|^3}
\end{align}
where $I_{c0}$ is the zero temperature current. For a cosine dispersion relation, $\disp{k} = \mu - J \sum_{i=x,y,z} \cos(k_i a)$, where $J$ represents the bandwidth, at half-filling ($\mu = 0$) the prefactor is given by $E^2 = \frac{3J^2}{2}$. Thus, at zero temperature, the ratio of the second-order current correction to the exact flat-band current is 
\begin{equation}
    \frac{I_0^{(2)}}{I_0^f} = \frac{9 J^2}{4 |\Delta|^2}
\end{equation}
In other words, the flat-band expression for the critical supercurrent, Eq.~(4) in the main text, is valid when $|\Delta| \gg J$. 

The temperature-dependent superconducting gap $\Delta(T)$, used in Fig.~2(b) of the main text, is solved self-consistently with BCS theory. In the dispersive case, the result is the usual BCS superconducting gap, with $\Delta(0)\approx 1.76 k_B T_c$ \cite{bardeen1957} and in the flat-band case the $\Delta(T)$ is solved from $2/U = \sum_{\vec{k}} \tanh\left(\sqrt{\Delta^2+ \disp{k}^2}/2k_B T\right) / \sqrt{\Delta^2+ \disp{k}^2}$, with $\disp{k}=0$, for which $\Delta(0) \approx 2 k_B T_c$.

\section{Tunnel junctions on a lattice}

The tunneling approximation for the supercurrent between two flat-band 
superconductors has an unusual feature: the result diverges as $1/\Delta$ when one sets $T\sim\Delta$. Here we derive the simple models used in the main text for understanding the result nonperturbatively.

Consider the BdG tight-binding Hamiltonian of a bilayer system, with weak tunneling
$T$ between the layers:
\begin{align}
  H
  =
  \begin{pmatrix}
    H_L & T^\dagger
    \\
    T & H_R
  \end{pmatrix}
  \,,
\end{align}
Consider now a point contact, where the tunneling amplitude matrix
element is nonzero only at a single position
\begin{align}
  T_{\vec{r}\vec{r}'}
  &=
  t \tau_3 \delta_{\vec{r},\vec{0}}\delta_{\vec{r}',\vec{0}}
  \,.
\end{align}
Here, $\vec{r}$ are unit cell coordinates, and $t=t^*$ is a matrix with a possible time-reversal symmetric orbital structure inside the unit cell.
With the usual Fourier definitions,
$\psi(\vec{k})=\sum_{\vec{r}}e^{i\vec{r}\cdot\vec{k}}\psi(\vec{r})\equiv\sum_{j}e^{i\vec{r}_j\cdot\vec{k}}\psi(\vec{r})$
sum over unit cells, and
$\psi(\vec{r})=\sum_{\vec{k}}e^{-i\vec{r}\cdot\vec{k}}\psi(\vec{k})\equiv\frac{A_{\rm uc}}{(2\pi)^d}\int_{\rm BZ}\dd{^dk}e^{-i\vec{r}\cdot\vec{k}}\psi(\vec{k})$, we have
$T_{\vec{k}\vec{k}'}=\sum_{\vec{r}\vec{r}'}e^{i\vec{r}\cdot\vec{k}-i\vec{r}'\cdot\vec{k}'}T_{\vec{r}\vec{r}'}=t\tau_3$. Here $A_{\rm uc}$ is the area of the unit cell.

We can recall the solution to this problem.  \cite{martinrodero1994}  The equation
$(\epsilon-H)G=1$ reads
\begin{align}
  &
  \begin{pmatrix}
    G_{L0,\vec{k}}^{-1} & 0
    \\
    0 & G_{R0,\vec{k}}^{-1}
  \end{pmatrix}
  G_{\vec{k},\vec{k}'}
  -
  \begin{pmatrix}
    0 & t^\dagger\tau_3
    \\
    t\tau_3 & 0
  \end{pmatrix}
  \sum_{\vec{q}}
  G_{\vec{q},\vec{k}'}
  =
  \delta_{\vec{k}\vec{k'}}
  \,,
\end{align}
where $G_{L/R0,\vec{k}}^{-1} = i\omega - H_{L/R}(\vec{k})$.
This results to
\begin{align}
  G_{\vec{k},\vec{k}'}
  &=
  \begin{pmatrix}
    0 &  G_{L0,\vec{k}} t^\dagger\tau_3
    \\
    G_{R0,\vec{k}} t\tau_3 & 0
  \end{pmatrix}
  \sum_{\vec{q}}
  G_{\vec{q},\vec{k}'}
  \\\notag&\quad
  +
  \begin{pmatrix}
    G_{L0,\vec{k}} & 0
    \\
    0 & G_{R0,\vec{k}}
  \end{pmatrix}
  \delta_{\vec{k}\vec{k'}}
  \,.
\end{align}
Summing over $\vec{k}$, and writing $g_{L/R}=\sum_{\vec{k}}G_{L/R0,\vec{k}}$,
\begin{align}
  \sum_{\vec{k}}G_{\vec{k},\vec{k}'}
  &=
  \begin{pmatrix}
    0 & g_{L} t^\dagger\tau_3
    \\
    g_{R} t\tau_3 & 0
  \end{pmatrix}
  \sum_{\vec{k}}
  G_{\vec{k},\vec{k}'}
  +
  \begin{pmatrix}
    G_{L0,\vec{k}'} & 0
    \\
    0 & G_{R0,\vec{k}'}
  \end{pmatrix}.
\end{align}
Hence,
\begin{align}
  G_{\vec{r}\vec{r}'=\vec{0}\vec{0}'}
  &=
  \sum_{\vec{k}\vec{k}'}G_{\vec{k},\vec{k}'}
  =
  \begin{pmatrix}
    1 & -g_{L} t^\dagger\tau_3
    \\
    -g_{R} t\tau_3 & 1
  \end{pmatrix}^{-1}
  \begin{pmatrix}
    g_{L} & 0
    \\
    0 & g_{R}
  \end{pmatrix}
\end{align}
The left/right off-diagonal blocks of this matrix read
\begin{align}
  G_{LR}
  &=
  (1 - g_L t^\dagger\tau_3 g_R t\tau_3)^{-1} g_L t^\dagger\tau_3 g_R
  \,,
  \\
  G_{RL}
  &=
  g_R t\tau_3 (1 - g_L t^\dagger \tau_3 g_R t\tau_3)^{-1} g_L
  \,.
\end{align}
The supercurrent between the layers is obtained by $t\mapsto{}te^{i\delta\varphi\tau_3/2}$
and varying the electronic free energy $F=T\sum_{\omega_n}\tr[i\omega_n - H]$ in $\delta\varphi$,
which gives the standard formula \cite{martinrodero1994}:
\begin{align}
  I_S
  &=
  \frac{e}{i \hbar}T\sum_{\omega_n}
  \tr( G_{LR} t - G_{RL} t^\dagger )
  \\
  \label{eq:ISextAB}
  &=
  \frac{e}{i \hbar}T\sum_{\omega_n}
  \tr\{(1 - g_L t^\dagger\tau_3 g_R t\tau_3)^{-1} g_L t^\dagger [\tau_3,g_R] t \}
  \\
  \label{eq:generalIS}
  &=
  \frac{-2e}{\hbar}T\sum_{\omega_n}
  \partial_{\delta\varphi}
  \ln \det D(\delta\varphi,\omega_n)\rvert_{\delta\varphi=0}
  \,,
  \\
  &D(\delta\varphi,\omega)
  =
  1 - g_L t^\dagger e^{-i\delta\varphi\tau_3/2} \tau_3 g_R t e^{i\delta\varphi\tau_3/2} \tau_3
  \,.
\end{align}
For $t\to0$, this reduces
to the Ambegaokar-Baratoff formula in the approximation of momentum-independent
tunneling. To find the current for an extended junction area $A$ in the
$t\to0$ limit, one can multiply the above single-site contribution by
$A/A_{\rm uc}$, the number of unit cells in the junction area.
This effectively assumes the junction interface is rough and does not
conserve momentum.

In the opposite case of a smooth interface between the layers,
$T_{\vec{r}\vec{r}'}=t\tau_3\delta_{\vec{r}\vec{r}'}$, one finds
Eq.~\eqref{eq:ISextAB} but with $g_{L/R}\mapsto{}G_{L/R0,\vec{k}}$ and
$\sum_{\omega_n}\mapsto (A/A_{\rm uc})\sum_{\omega_n,\vec{k}}$.  For a flat band
filling the whole Brillouin zone, one has $g_{L/R}=G_{L/R0,\vec{k}}$, and
both cases give the same supercurrent. In contrast, for dispersive
bands whether the interface conserves parallel momentum or not can
matter.

With the simple flat-band model of the main text, Eq.~\eqref{eq:ISextAB}
for the single-site current becomes:
\begin{align}
  I_S
  =
  \frac{4e}{i\hbar}
  \sum_{\omega}
  \frac{
    T \Delta_S^2 |t|^2\sin\phi
  }{
    \prod_\pm (\omega_n^2 + \Delta_S^2 + |t|^2 \pm 2 \Delta_S |t| \sin\frac{\varphi}{2})
  }
  \,.
\end{align}
The terms $\propto|t|$ in the denominator provide a cutoff for the
$1/T$ divergence of the supercurrent.

Evaluating the Matsubara sum gives the supercurrent in the
Andreev bound state form:
\begin{align}
  I_S
  &=
  \frac{e}{2\hbar}
  \sum_\pm
  \frac{
    \mp\Delta_S |t| \sin(\varphi)
    \tanh\frac{\sqrt{\Delta_S^2 + |t|^2 \pm 2\Delta_S|t\sin\frac{\varphi}{2}|}}{2T}
  }{
    |\sin\frac{\varphi}{2}|\sqrt{\Delta_S^2 + |t|^2 \pm 2\Delta_S|t\sin\frac{\varphi}{2}|}
  }
  \\
  &=
  -
  \frac{2e}{\hbar}
  \sum_\pm
  \frac{\partial\epsilon_\pm}{\partial\varphi}
  \tanh\frac{\epsilon_\pm}{2T}
  \,,
  \label{eq:Isinglechannel}
  \\
  \epsilon_\pm
  &=
  \sqrt{\Delta_S^2 + |t|^2 \pm 2\Delta_S t\sin\frac{\varphi}{2}}
  \,.
\end{align}
For $|t|\to0$, this approaches the Ambegaokar-Baratoff result.

The energies $\epsilon_\pm$ are also the eigenenergies of the Hamiltonian
\begin{align}
  \label{eq:Htrivialstates}
  H =
  \begin{pmatrix}
    0 & \Delta_S & t & 0
    \\
    \Delta_S & 0 & 0 & -t
    \\
    t & 0 & 0 & \Delta_S e^{i\varphi}
    \\
    0 & -t & \Delta_S e^{-i\varphi} & 0
  \end{pmatrix}
  \,.
\end{align}
In this minimal model, the flat band is thought
to be formed by localized quasiparticle states, between which the
tunneling occurs. The tunneling is then always a local process, and
momentum conservation at the interface does not matter, which is why
the extended interface model, the point-contact model, and this
two-state model, give equivalent results for the supercurrent per
tunneling site.

Note that the above result differs from the usual dispersive superconductor
point contact supercurrent, \cite{beenakker1991universal} which is generated by an Andreev bound state with
energy $\epsilon_A=\Delta\sqrt{1 - \tau \sin^2\frac{\varphi}{2}}$,
where $0\le\tau\le1$ is a junction transparency. The transparency
is a function of the tunneling amplitudes and the dispersion parameters
\cite{martinrodero1994}.

For more complex models, the result also depends on which sites in the
unit cell the tunneling matrix $t$ connects, as the currents may
interfere destructively.
For an isolated time-reversal symmetric flat band, Green's function
$G=(i\omega-H_{\mathrm{BdG}})^{-1}$ can be approximated by its flat-band part,
as other terms are suppressed by the band gaps. Then, in Eq.~\eqref{eq:generalIS}
\begin{gather}
  g_{L/R}
  \approx
  g_{0,L/R}
  \otimes
  P_0
  \,,
  \quad
  P_0
  =
  \sum_{\vec{k}}
  \psi_{\vec{k},0}\psi_{\vec{k},0}^\dagger
  \,,
  \\
  g_{0,L/R}
  =
  \frac{-1}{\omega^2 + \varepsilon_0^2 + |\tilde{\Delta}|^2}
  \begin{pmatrix}
    i\omega + \varepsilon_0 & \tilde{\Delta} \\
    \tilde{\Delta}^* & i\omega - \varepsilon_0
  \end{pmatrix}
  \,,
\end{gather}
where $\varepsilon_0$ and $\psi_{\vec{k},0}$ are the energy and Bloch
functions of the flat band and $\tilde{\Delta}$ its superconducting
gap. Then, $\det D = \prod_j \det(1 - \lambda_j
g_{0,L}\tau_3e^{-i\delta\varphi\tau_3/2}g_{0,R}\tau_3e^{i\delta\varphi\tau_3/2})$,
where $\lambda_j$ are the eigenvalues of $P_{0L} t^\dagger P_{0R} t$. The
total supercurrent is then found by summing the single-channel result~\eqref{eq:Isinglechannel}
over the effective tunnel amplitudes $|t|^2\mapsto|\lambda_j|$.
The corresponding Andreev bound states and supercurrent are
\begin{subequations}
  \label{eq:ABSeff}
\begin{align}
  \label{eq:ABSeffen}
  \epsilon_{j\pm}
  &=
  \sqrt{\Delta_S^2 + |\lambda_j| \pm 2\Delta_S \sqrt{|\lambda_j|} \sin\frac{\varphi}{2}}
  \,,
  \\
  I_S
  &=
  -
  \frac{2e}{\hbar}
  \sum_{j,\pm}
  \frac{\partial\epsilon_{j,\pm}}{\partial\varphi}
  \tanh\frac{\epsilon_{j,\pm}}{2T}
\end{align}
\end{subequations}
as determined by the amplitudes $\lambda_j$.

\begin{figure}
  \includegraphics{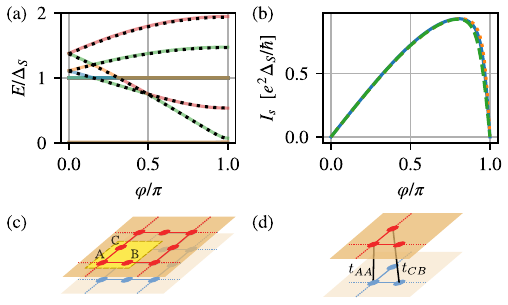}
  \caption{\label{fig:ABSeff}
    (a)
    Bound states in a point contact between two 2D Lieb lattices,
    with hopping $t_0=50$ and gap $\Delta_S=1$, as a function of the superconducting
    phase difference $\varphi$ between the layers.
    Solid: eigenstates in numerical diagonalization of two $25\times{}25$ unit cell lattices connected by point contact.
    Dotted: Eq.~\eqref{eq:ABSeff}.
    (b)
    Corresponding supercurrent at zero temperature.
    Solid: finite-size lattice.
    Dashed: infinite lattice.
    Dotted: Eq.~\eqref{eq:ABSeff}.
    (c)
    Schematic of the Lieb lattice bilayer, with the 2D unit cell
    and site labels indicated.
    (d)
    Schematic of a point contact between the layers. The two layers are connected
    by a tunneling matrix $t_{\alpha\beta}$ in a single unit cell of the lattice.
  }
\end{figure}

An example of the results Eq.~\eqref{eq:ABSeff} and their comparison
to numerical studies is shown in Fig.~\ref{fig:ABSeff}, in the case of
two 2D Lieb lattices \cite{lieb1989} coupled by a point contact.  A
schematic of the lattice model is shown in Fig.~\ref{fig:ABSeff}(c):
in both layers, there are three orbitals A, B, C in the 2D unit
cells. The 2D lattice hosts a flat band at $E=0$, and two dispersive
Dirac cone-like bands.  We assume the layers have superconducting
pair potential $\Delta=(\Delta_A,\Delta_B,\Delta_C)$ on the orbitals,
with $\Delta_A\approx0$ and $\Delta_B=\Delta_C\equiv\Delta_S$,
corresponding to the orbital structure of the flat band. The point
contact connects the two layers together as shown in
Fig.~\ref{fig:ABSeff}(d) at a single position (unit cell) in the
lattice.  Its orbital structure is specified by the point contact
interlayer hopping matrix $t_{\alpha\beta}$, where
$\alpha,\beta\in\{A,B,C\}$. The results shown are calculated taking
hopping $t_0=50\Delta_S$ for the lattice.  The choice of the point
contact hopping matrix $t$ changes the effective amplitudes
$|\lambda_j|$.  To give an example, the results are calculated with
\begin{align}
  t
  =
  \begin{pmatrix}
    0.5 & 0 & 1 \\
    0.5 & 2 & 0 \\
    0   & 1 & 1
  \end{pmatrix}
  \Delta_S
  \,.
\end{align}
The eigenenergies of the bilayer BdG Hamiltonian of a finite-size
lattice are shown in Fig.~\ref{fig:ABSeff}(a), together with
Eq.~\eqref{eq:ABSeffen} with $\lambda_j$ corresponding to the chosen
$t$. The agreement is good: mostly the flat band at Fermi level
contributes to the phase-dependent bound states.  The Dirac cones of
the continuum model appear to contribute little: their density of
states is low compared to the flat band, and in addition in this
finite-size calculation the size quantization is very large so that
their contribution is gapped.

Calculating the supercurrent in Fig.~\ref{fig:ABSeff}(b), we see
Eq.~\eqref{eq:ABSeff} is accurate also for it.  Here, using
Eq.~\eqref{eq:generalIS} we can moreover obtain the point-contact
supercurrent in an infinite lattice without size quantization issues.
There turns out to be little difference to the finite-size result of
Fig.~\ref{fig:ABSeff}(a) and to Eq.~\eqref{eq:ABSeff}.  This is
because the point contact probes states locally, and the dominant
flat-band contribution comes by nature from localized BdG states and does
not depend on the lattice size.

\section{Compactly localized evanescent modes}

To understand how the S/N Andreev reflection problem~\cite{blonder1982-tfm} discussed in the main text
behaves when the superconductor is a flat-band system, 
and how this is visible in the resulting scattering amplitude formulas,
it is useful to first revisit the textbook calculation.
Below, we first point out the main physics and then describe
the lattice formulation of the scattering problem and its general solution.

Consider an S/N interface with a superconductor at $x<0$ and a normal metal at $x>0$.
In the textbook Andreev reflection calculation, evanescent modes at $x<0$ are matched to propagating modes at $x>0$, which determines the scattering amplitudes.

Assume now the superconductor has an exact flat band with the energy dispersion
\begin{align}
  \epsilon_{\vec{k}n} = \epsilon_n = \mathrm{constant}
  .
  \label{eq:fb}
\end{align}
Evanescent modes $\psi(\vec{r})\propto{}e^{\kappa(\epsilon) x}$, or
in lattice models $\psi_j\propto{}e^{\kappa(\epsilon) a j}$ with $j$ the unit
cell index, are (unbounded) solutions of the BdG equation at
energy $\epsilon$ inside the energy gap of S.
The wave vectors $\kappa=ik_x$ are determined by
the complex roots of $\epsilon_{\vec{k}n}=\epsilon$. However, the
analytic continuation of a constant to the complex plane is a
constant. This implies that the continuum formulation of the problem
is somewhat ill-defined: no such modes exist at energies
$\epsilon\ne\epsilon_n$, reflecting the zero group velocity and
localization of the quasiparticle states in a flat band.
Conversely, exactly at $\epsilon = \epsilon_n$, 
the equation would in principle be solved by any $\vec{k}$, which
reflects the existence of a large number of states at the exact flat band energy.

When there is slight dispersion,
\begin{align}
  \epsilon_{\vec{k}n} = \epsilon_n + \eta g(\vec{k})
  ,
  \label{eq:sldisp}
\end{align}
where $\eta$ is small, in the ``usual'' case (e.g., analytic $g$) the
equation $\epsilon_{\vec{k}n}=\epsilon$ has solutions
$\vec{k}=g^{-1}(\frac{\epsilon-\epsilon_n}{\eta})$ for almost all
$\epsilon$.  Also, as $\eta\to0$ the evanescent modes would tend
toward large $|\vec{k}|$, and their decay becomes faster as the band
flattens. In the exact flat band limit, the evanescent decay factor
$\lambda=e^{\kappa a}$ must then be $\lambda=0$ (or $\lambda=\infty$
for waves decaying to opposite direction); that is, there are no \emph{finite}
solutions the equation $\epsilon_{\vec{k}n}=\epsilon$.  In lattice models, this
however does not imply that evanescent waves inside the flat band drop
to zero within one unit cell, as the above only rules out exponential
solutions. A possibility that is left is nilpotent decay, where the
wave function becomes zero after some number of unit cells.

Consider now a noninteracting (quasi) one-dimensional tight-binding model,
with the block structure
\begin{align}
  \hat{H}_{i,j}
  =
  \delta_{i-1,j} \hat{J}_j + \delta_{i,j} \hat{K}_j + \delta_{i+1,j} \hat{J}_{j+1}^\dagger
  .
\end{align}
Here $\hat{J}$ of size $N\times{}N$ describes the hopping between unit
cells with $N$ sites. Choosing the unit cell large enough, 
all models with finite-range hopping
can be brought to this form. Let us first assume $\hat{J}_j$ and
$\hat{K}_j$ are independent of $j$.

The evanescent modes at energy $\epsilon$ are local solutions to the equation
$(\hat{H}-\epsilon)\psi=0$.  It is convenient to rewrite the problem
as a transfer matrix equation~\cite{lee1981}, defining $\phi_j=(\psi_j; \psi_{j+1})$,
\begin{align}
  \label{eq:tmeq}
  A\phi_{j-1}
  &\equiv
  \begin{pmatrix}
    J & K-\epsilon
    \\
    0 & 1
  \end{pmatrix}
  \phi_{j-1}
  =
  \begin{pmatrix}
    0 & -J^\dagger
    \\
    1 & 0
  \end{pmatrix}
  \phi_j
  \equiv
  B\phi_j
  .
\end{align}
The transfer matrix equation for the evanescent modes is associated
with the generalized eigenproblem for $A-\lambda{}B$. In the presence
of flat bands, the eigenproblem is not necessarily diagonalizable, as
the generalized eigenvalues $0,\infty$ may be highly degenerate and do
not originate simply from the zero modes of the hopping matrix.
Moreover, transfer matrices $T=A^{-1}B$ or $T=B^{-1}A$ are
ill-defined; a system with exact flat bands has only generalized
eigenvalues $0,\infty$ and the situation cannot be corrected by making
basis transformations \cite{dwivedi2016-bbg}.

Mathematically, the problem is well understood. The structure is
best visible by transforming to the Kronecker canonical form~\cite{gantmakher1959-tm,ikramov1993-mpt}, %
$A=U\tilde{A}V^{-1}$, $B=U\tilde{B}V^{-1}$.
We consider now evanescent modes and assume $\epsilon$ does not coincide with the spectrum of the Hamiltonian.
In this case, the eigenproblem is \emph{regular}.
Then, $\tilde{A}$, $\tilde{B}$ are block-diagonal with blocks of size $m_k$ for each
generalized eigenvalue $\lambda_k$.

We have $\det(\mu A - \lambda B) = \det(\mu^2 J +
(K-\epsilon)\mu\lambda + \lambda^2 J^{\dagger})=\mu^N\lambda^N\det[H(k=-i\ln\frac{\lambda}{\mu})-\epsilon]$
which is not identically zero except possibly at special values of
$\epsilon$.  Hence, the matrix pencil $A-\lambda{}B$ is generically regular and
its decomposition contains only the Weierstrass part.

Defining
$\tilde{\phi}_j=V^{-1}\phi_j$, Eq.~\eqref{eq:tmeq} separates to blocks of two types
\begin{align}
  \label{eq:weierstrass}
  J_{m_k}(\lambda_k) \tilde{\phi}_{j-1}^k = \tilde{\phi}_j^k
  ,
  \quad\text{or}\quad
  \tilde{\phi}_{j-1}^k = J_{m_k}(0) \tilde{\phi}_j^k
  ,
\end{align}
where $\tilde{\phi}_j=(\tilde{\phi}^1_j; \tilde{\phi}^2_j; \ldots;
\tilde{\phi}^M_j)$, $\sum_{k=1}^M m_k=2N$, and $[J_{m}(\lambda)]_{ij}
= \lambda\delta_{ij} + \delta_{j,i+1}$ are Jordan blocks of size
$m\times{}m$ for the eigenvalues $\lambda_k$. Finite eigenvalues have
blocks of the first type, and infinite eigenvalues $\lambda=\infty$
produce blocks of the second type. The same eigenvalue may have multiple
blocks.  The blocks with $|\lambda_k|>1$ describe modes decaying
toward $j\to-\infty$, $|\lambda_k|<1$ those decaying for
$j\to+\infty$, and $|\lambda_k|=1$ the propagating modes except for
$m_k>1$ of those $m_k-1$ are polynomially growing.

Pure flat-band systems at $\epsilon\ne\epsilon_n$ have only
eigenvalues $\lambda_k=0,\infty$. Indeed, $\det(\mu A-\lambda
B)=\mu^N\lambda^N\prod_n(\epsilon_n-\epsilon)$, so both have
multiplicity $N$. Moreover, the coefficient matrix of
Eqs.~\eqref{eq:weierstrass} is nilpotent, $J_m(0)^m=0$, and therefore
these evanescent modes become zero after $m_k\le{}N$ lattice sites.
This describes how the localized states in the flat band system
participate in evanescent transport.  If all $m_k=1$ for a pure
flat-band system, then $J=0$, so nontrivial pure flat-band systems
generally have some $m_k>1$.

Numerically, the Kronecker decomposition is unstable to calculate, and
one should instead use the generalized Schur decomposition. Schur decomposition
algorithms for solving the quantum transport problem
\cite{wimmer2008-qtn} generally work also for the
flat-band lattices.  Analytically, the Kronecker decomposition however
tends to be easier to calculate.

The transition in the wave functions when going from slight dispersion
Eq.~\eqref{eq:sldisp} to exact flat band Eq.~\eqref{eq:fb} should be
smooth. However, the nontrivial Jordan block structure with $m_k>1$ is
fragile and requires fine-tuning conditions that can exist in exact
flat bands, but would not be present in generically perturbed slightly
dispersive models.  That there is no contradiction is a consequence of
the non-Hermiticity of the evanescent mode problem.  As the dispersion
approaches a flat band with some $m_k>1$, the eigenvectors (columns of
$V$) of corresponding groups of eigenvalues become increasingly parallel. This enables approaching the
nilpotent decay as $\eta\to0$ despite $|\vec{k}|\to\infty$, and
ensures that the exact flat band limit can give useful insight also
into models where the exact flat band condition is lifted by slight dispersion.

\section{Obtaining the normal state Green's function $g$ from the Kronecker decomposition}

In the problems of finding the supercurrent or Andreev reflection amplitude
discussed in the main text, the necessary information about a semi-infinite lead
is captured by the boundary Green's function~\cite{datta1995}.
It can be obtained from the transfer matrix formulation as follows.

Consider now the boundary condition in the semi-infinite lead case,
$J_j=J$, $K_j=K$ for $j\le0$. The transfer matrix equation at $j=0$
is
\begin{align}
  A \phi_{-1} = \begin{pmatrix} 0 & -J_1^\dagger \\ 1 & 0 \end{pmatrix} \begin{pmatrix} \psi_0 \\ \psi_1 \end{pmatrix}
  .
\end{align}
Using the semi-infinite lead boundary condition
$\tilde{\phi}^k_{-1}=0$ for $|\lambda_k|<1$ (modes growing toward left), gives then
\begin{align}
  W_2 \psi_0 - W_1 J_1^\dagger \psi_1
  &=
  0
  ,
  &
  \begin{pmatrix} W_1 & W_2 \end{pmatrix}
  &=
  P_< U^{-1}
  \,,
\end{align}
where the matrix $U$ originates from the Kronecker decomposition, and
$P_<$ has rows containing zeros and a single $1$ selecting each growing mode.
For generic complex
$\epsilon$, the transfer matrix equation has generally $N$ such
modes. If the $|\lambda_k|<1$
eigenvalues are sorted first in the Kronecker decomposition, then
$P_<=\begin{pmatrix}1_{N\times{}N} & 0_{N\times{}N}\end{pmatrix}$.
Then also $W_{1,2}$ are of size $N\times{}N$, and we can rewrite
this equation as
\begin{align}
  \label{eq:bc-eqn-exp}
  -g^{-1} \psi_0 + J_1^\dagger \psi_1 &= 0
  ,
  &
  g
  &=
  W_2^{-1} W_1
  \,.
\end{align}
The lead lattice sites $j<0$ are now eliminated from the equation
$(H-\epsilon)\psi=0$, and replaced by this boundary condition at
$j=0$. From this equation, we can identify $g$ as the surface Green's
function of the lead. Although not easily seen in this representation, it satisfies
the recurrence equation $g = (\epsilon - K - J g J^\dagger)^{-1}$.

The above calculation gives $g$ in terms of the Kronecker
decomposition.  A similar result can be found for the Schur
decomposition \cite{wimmer2008-qtn}.

\section{Scattering problem}

Let us now derive the expression for the backscattering matrix given
in the main text. The normal state part of the system is on the right and the superconductor on the left.

Consider a simple 1D chain normal lead in the BdG problem, $K=0$, $J=t_N\tau_3$,
where $t_N$ is the hopping of the normal-state 1D chain.
The transfer matrix can be diagonalized, $T=B^{-1}A=V \Lambda V^{-1}$.
The left and right propagating blocks are
\begin{align}
  V_{\leftarrow/\rightarrow} &=
  \begin{pmatrix}
    1 \\ \Lambda_{\leftarrow/\rightarrow}
  \end{pmatrix}
  \,,
  &
  \Lambda_{\leftarrow/\rightarrow} &= \tau_3 e^{\mp{}ik_e}
  \,,
  \\
  \Lambda
  &=
  \diag(\Lambda_\leftarrow,\Lambda_\rightarrow)
  \,,
\end{align}
where $k_e=\pi-k_h=\arccos(\epsilon/2t_N)$ with branch choice $0\le{}k_e\le\pi$.
The group velocity is $v_{ge}=-t_n\sin{}k_e=-v_{gh}$.

The scattering modes at $j\ge0$ are then
\begin{align}
  \phi_j
  &=
  V_{\leftarrow} \Lambda_\leftarrow^j a_{\rm in} + V_{\rightarrow} \Lambda_\rightarrow^j a_{\rm out}
  \,,
\end{align}
where $a_{\rm in}=(a_e; a_h)$ and $a_{\rm out}=(b_e; b_h)$ are
$4\times1$ vectors containing the incoming and outgoing electron/hole amplitudes.
Here, $\phi_j=(\psi_j,\psi_{j+1})$ is the vector appearing in the transfer
matrix problem above.

The backscattering matrix can then be directly solved from the boundary
condition \eqref{eq:bc-eqn-exp}:
\begin{align}
  R
  =
  -
  \frac{
    1 - g t_N e^{-ik_e}
  }{
    1 - g t_N e^{ik_e}
  }
  \,.
\end{align}
Note that in this formulation, $R$ relates the incoming and outgoing modes
on the right to each other,
\begin{align}
   a_{\rm out} = R a_{\rm in}
   \,.
\end{align}
The propagating modes (if any) in the superconductor on the left, are contained in $g$. For a fully
reflecting boundary, $g=g^\dagger$ and $R$ is unitary.  When the
boundary is partially transmitting, $R$ is not unitary, and the
``lost'' probability current corresponds to probability current
flowing into the superconductor.

\section{Creutz ladder flat band}

To find the boundary Green's function $g$ for the Creutz ladder needed in the Andreev reflection problem,
and to illustrate the Kronecker decomposition, below we solve the boundary
problem for the Creutz ladder analytically, in its flat-band limit. Again, the superconductor is assumed to be in the left and the normal state system on the right.

Consider the BdG model in the Creutz ladder, for the
flat-band case
\begin{align}
  \label{eq:creutz-J}
  J =
  \tau_3
  \otimes
  \begin{pmatrix}
    t & -t
    \\
    t & -t
  \end{pmatrix}
  ,
  \qquad
  K
  &=
  \Delta\tau_1 - \mu
  \,,
\end{align}
where $\tau_{1,2,3}$ are the Pauli matrices in the Nambu space.
The transfer matrix equation can be written with
$A=u^\dagger A' u$,  $B=u^\dagger B' u$, where
\begin{align}
  A'
  &=
  \begin{pmatrix}
    0 & 2 \, t & 0 & 0 & -\epsilon - \mu & 0 & \Delta & 0 \\
    0 & 0 & 0 & 0 & 0 & -\epsilon - \mu & 0 & \Delta \\
    0 & 0 & 0 & -2 \, t & \Delta & 0 & -\epsilon + \mu & 0 \\
    0 & 0 & 0 & 0 & 0 & \Delta & 0 & -\epsilon + \mu \\
    0 & 0 & 0 & 0 & 1 & 0 & 0 & 0 \\
    0 & 0 & 0 & 0 & 0 & 1 & 0 & 0 \\
    0 & 0 & 0 & 0 & 0 & 0 & 1 & 0 \\
    0 & 0 & 0 & 0 & 0 & 0 & 0 & 1
  \end{pmatrix}
  \,,
\end{align}
\begin{align}
  B'
  &=
  \begin{pmatrix}
    0 & 0 & 0 & 0 & 0 & 0 & 0 & 0 \\
    0 & 0 & 0 & 0 & -2 \, t & 0 & 0 & 0 \\
    0 & 0 & 0 & 0 & 0 & 0 & 0 & 0 \\
    0 & 0 & 0 & 0 & 0 & 0 & 2 \, t & 0 \\
    1 & 0 & 0 & 0 & 0 & 0 & 0 & 0 \\
    0 & 1 & 0 & 0 & 0 & 0 & 0 & 0 \\
    0 & 0 & 1 & 0 & 0 & 0 & 0 & 0 \\
    0 & 0 & 0 & 1 & 0 & 0 & 0 & 0
  \end{pmatrix}
  \,,
  \\
  u
  &=
  \frac{1}{\sqrt{2}}
  1_{4\times4}\otimes\begin{pmatrix}1&1\\1&-1\end{pmatrix}
  \,.
\end{align}
The Kronecker decomposition of $(A', B')$ is
\begin{align}
  a'
  &=
  \begin{pmatrix}
    0 & 1 & 0 & 0 & 0 & 0 & 0 & 0 \\
    0 & 0 & 0 & 0 & 0 & 0 & 0 & 0 \\
    0 & 0 & 0 & 1 & 0 & 0 & 0 & 0 \\
    0 & 0 & 0 & 0 & 0 & 0 & 0 & 0 \\
    0 & 0 & 0 & 0 & 1 & 0 & 0 & 0 \\
    0 & 0 & 0 & 0 & 0 & 1 & 0 & 0 \\
    0 & 0 & 0 & 0 & 0 & 0 & 1 & 0 \\
    0 & 0 & 0 & 0 & 0 & 0 & 0 & 1
  \end{pmatrix}
  \,,
  \\
  b'
  &=
  \begin{pmatrix}
    1 & 0 & 0 & 0 & 0 & 0 & 0 & 0 \\
    0 & 1 & 0 & 0 & 0 & 0 & 0 & 0 \\
    0 & 0 & 1 & 0 & 0 & 0 & 0 & 0 \\
    0 & 0 & 0 & 1 & 0 & 0 & 0 & 0 \\
    0 & 0 & 0 & 0 & 0 & 1 & 0 & 0 \\
    0 & 0 & 0 & 0 & 0 & 0 & 0 & 0 \\
    0 & 0 & 0 & 0 & 0 & 0 & 0 & 1 \\
    0 & 0 & 0 & 0 & 0 & 0 & 0 & 0
  \end{pmatrix}
  \,.
\end{align}
There are two $2\times2$ Jordan blocks corresponding to the
generalized eigenvalue 0 with degeneracy 4 (sorted first in the
decomposition), and similarly for the infinite eigenvalue. The size of
these blocks directly determines the compact localization of the
evanescent states.

Note that the decomposition is not valid at the flat-band energies,
$\epsilon=\pm\sqrt{\Delta^2 + (\mu \pm 2t)^2}$, where the generalized
eigenvalue problem is no longer regular. In this case, the
decomposition contains non-square blocks that correspond to the
compactly localized flat-band eigenstates.

The transformation matrices are (via computer algebra, see
\texttt{creutz.sage} \cite{codes})
\begin{widetext}
\begin{align}
  U'
  &=
  \begin{pmatrix}
0 & 0 & 0 & 0 & 0 & -\frac{\Delta^{2} \epsilon - \epsilon^{3} + \Delta^{2} \mu - \epsilon^{2} \mu + \epsilon \mu^{2} + \mu^{3} + 4 \, \epsilon t^{2} - 4 \, \mu t^{2}}{\Delta^{2} - \epsilon^{2} + \mu^{2}} & 0 & \frac{{\left(\Delta^{2} - \epsilon^{2} + \mu^{2} + 4 \, t^{2}\right)} \Delta}{\Delta^{2} - \epsilon^{2} + \mu^{2}} \\
0 & -2 \, t & 0 & 0 & -2 \, t & 0 & 0 & 0 \\
0 & 0 & 0 & 0 & 0 & \frac{{\left(\Delta^{2} - \epsilon^{2} + \mu^{2} + 4 \, t^{2}\right)} \Delta}{\Delta^{2} - \epsilon^{2} + \mu^{2}} & 0 & -\frac{\Delta^{2} \epsilon - \epsilon^{3} - \Delta^{2} \mu + \epsilon^{2} \mu + \epsilon \mu^{2} - \mu^{3} + 4 \, \epsilon t^{2} + 4 \, \mu t^{2}}{\Delta^{2} - \epsilon^{2} + \mu^{2}} \\
0 & 0 & 0 & 2 \, t & 0 & 0 & 2 \, t & 0 \\
1 & 0 & 0 & 0 & 0 & 1 & 0 & 0 \\
0 & \frac{\epsilon + \mu}{2 \, t} & 0 & -\frac{\Delta}{2 \, t} & -\frac{2 \, {\left(\epsilon - \mu\right)} t}{\Delta^{2} - \epsilon^{2} + \mu^{2}} & 0 & \frac{2 \, \Delta t}{\Delta^{2} - \epsilon^{2} + \mu^{2}} & 0 \\
0 & 0 & 1 & 0 & 0 & 0 & 0 & 1 \\
0 & \frac{\Delta}{2 \, t} & 0 & -\frac{\epsilon - \mu}{2 \, t} & -\frac{2 \, \Delta t}{\Delta^{2} - \epsilon^{2} + \mu^{2}} & 0 & \frac{2 \, {\left(\epsilon + \mu\right)} t}{\Delta^{2} - \epsilon^{2} + \mu^{2}} & 0
  \end{pmatrix}
  \,,
  \\
  V'
  &=
  \begin{pmatrix}
1 & 0 & 0 & 0 & 0 & 0 & 0 & 0 \\
0 & \frac{\epsilon + \mu}{2 \, t} & 0 & -\frac{\Delta}{2 \, t} & 0 & -\frac{2 \, {\left(\epsilon - \mu\right)} t}{\Delta^{2} - \epsilon^{2} + \mu^{2}} & 0 & \frac{2 \, \Delta t}{\Delta^{2} - \epsilon^{2} + \mu^{2}} \\
0 & 0 & 1 & 0 & 0 & 0 & 0 & 0 \\
0 & \frac{\Delta}{2 \, t} & 0 & -\frac{\epsilon - \mu}{2 \, t} & 0 & -\frac{2 \, \Delta t}{\Delta^{2} - \epsilon^{2} + \mu^{2}} & 0 & \frac{2 \, {\left(\epsilon + \mu\right)} t}{\Delta^{2} - \epsilon^{2} + \mu^{2}} \\
0 & 1 & 0 & 0 & 0 & 1 & 0 & 0 \\
0 & 0 & 0 & 0 & -\frac{2 \, {\left(\epsilon - \mu\right)} t}{\Delta^{2} - \epsilon^{2} + \mu^{2}} & 0 & \frac{2 \, \Delta t}{\Delta^{2} - \epsilon^{2} + \mu^{2}} & 0 \\
0 & 0 & 0 & 1 & 0 & 0 & 0 & 1 \\
0 & 0 & 0 & 0 & -\frac{2 \, \Delta t}{\Delta^{2} - \epsilon^{2} + \mu^{2}} & 0 & \frac{2 \, {\left(\epsilon + \mu\right)} t}{\Delta^{2} - \epsilon^{2} + \mu^{2}} & 0
  \end{pmatrix}
  \,.
\end{align}
\end{widetext}
The decomposition of $(A, B)$ is $(UaV^{-1}, UbV^{-1})$ with $a=a'$, $b=b'$, $U=u^\dagger U'$, $V=u^\dagger V'$.

Knowing the decomposition, we can find the surface Green's function $g$ from
Eq.~\eqref{eq:bc-eqn-exp}. By direct calculation, we find a simple
result
\begin{align}
  g &= (\epsilon - K - \Sigma)^{-1}
  \,,
  \\
  \Sigma
  &=
  \frac{2t^2}{\Delta^2+\mu^2-\epsilon^2}
  \begin{pmatrix}
    \mu - \epsilon & \Delta
    \\
    \Delta & -\mu - \epsilon
  \end{pmatrix}
  \otimes
  \begin{pmatrix}
    1 & 1
    \\
    1 & 1
  \end{pmatrix}
  \,,
\end{align}
for the right edge of the semi-infinite chain.  The pole in $\Sigma$
is from the edge state of the Creutz ladder.

The surface Green's function itself then is
\begin{align}
  g_\gamma
  &=
  \frac{1}{2}
  \frac{1}{\Delta^2 + \mu^2 - \epsilon^2}
  \begin{pmatrix}
    \mu - \epsilon & -\Delta
    \\
    -\Delta & -\mu - \epsilon
  \end{pmatrix}
  \otimes
  \begin{pmatrix}
    1 & -\gamma \\ -\gamma & 1
  \end{pmatrix}
  \notag
  \\
  &+\frac{1}{4}
  \sum_{\alpha=\pm}
  \frac{1}{\Delta^2 + (\mu - 2\alpha t)^2 - \epsilon^2}
  \notag
  \\&\qquad\times
  \begin{pmatrix}
    \mu - 2\alpha t - \epsilon & -\Delta
    \\
    -\Delta & -\mu + 2\alpha t - \epsilon
  \end{pmatrix}
  \otimes
  \begin{pmatrix}
    1 & \gamma \\ \gamma & 1
  \end{pmatrix}
  \,,
\end{align}
where $\gamma=+1$ for the right edge of a semi-infinite chain and
$\gamma=-1$ for the left edge, which can be found from a similar
calculation. Here, $\alpha=\pm$ is the band index.
The first line is the edge state contribution and the
remainder is that of the flat bands at $\epsilon=\pm2t$.

The suppression of tunneling current shown in Fig.~3 of the main text is
also apparent here:
\begin{align}
  I_c
  \propto \tr f_R J_T f_L J_T^\dagger
  \,.
\end{align}
Here $f_L$ and $f_R$ are the anomalous Green's functions on the left/right sides
of the tunnel junction, and $J_T$ is the tunneling amplitude matrix.
If $J_T\propto{}J$, then $J_T f_L J_T^\dagger\propto\Sigma$.  Since
$\Sigma$ contains only the pole contribution from the edge state which
is separated in energy from the flat band by $t$, the current is
suppressed by a factor of $(\Delta/t)^2$ when the chemical potential
$\mu=\pm 2t$ is aligned with the flat band.

This behavior originates from the recurrence relation $(\epsilon - K
- J g J^\dagger)g = 1$ of the surface Green's function, and appears fairly
generic across quasi-1D exact flat-band lattice models.
Namely, separating a flat-band pole part $g = g_{\rm pole} + g_{\rm smooth}$,
$g_{\rm pole} = A_{\rm pole} / (\epsilon - \epsilon_{FB})$, the relation becomes
\begin{align}
\notag
 J g_{\rm pole} J^\dagger g_{\rm pole} 
 &=
 (\epsilon - K)g - 1 
 - J g_{\rm smooth} J^\dagger g_{\rm smooth}
 \\&
 - J g_{\rm smooth} J^\dagger g_{\rm pole}
 - J g_{\rm pole} J^\dagger g_{\rm smooth}
 \,,
\end{align}
where the right-hand side contains at most one $g_{\rm pole}$ in each
term. Collecting terms in different orders in $\propto{}(\epsilon
- \epsilon_{\rm FB})^{-1}$ in the limit $\epsilon\to\epsilon_{FB}$ one
finds that $J A_{\rm pole} J^\dagger A_{\rm pole} = 0$. This then
can lead to cancellations in tunneling amplitudes.

Using the above results for $g$ in the expression for the backscattering
matrix $R$ produces the results show in Fig.~4(a) of the main text.

\section{Interaction-mediated current}

We now calculate the critical current in the case the N-material has effective attractive interactions of strength $U$ between the electrons, using a Ginzburg-Landau approach. 
The Ginzburg--Landau expansion \cite{larkin2005} can be found by considering 
the mean-field free energy of electrons, where a mean field $\Delta$ describes
the attractive interactions, and expanding in small $\Delta$, namely,
\begin{align}
  F
  &=
  \sum_x \frac{|\Delta(x)|^2}{U}
  -
  T
  \sum_\omega \tr\ln[i\omega - H]
  \\
  &\simeq
  \sum_{xx'} V(x,x') \Delta(x)^* \Delta(x')
  \,.
\end{align}
Often the Ginzburg--Landau theory is done in gradient expansion, so that
\begin{align}
  F
  &=
  \sum_x [\mathcal{A}\frac{D}{2} |\partial_x \Delta(x)|^2 + \mathcal{A} b |\Delta(x)|^2]
  \\
  V(x,x') &= [-\mathcal{A}\frac{D}{2} \partial_x^2 + \mathcal{A} b] \delta(x-x')
  \,.
\end{align}
Above, we considered a quasi-1D problem, i.e., $\Delta(\vec{x})=\Delta(x)$.

Consider then a situation where $\Delta(x=0)$ is fixed, and other values
of $\Delta$ are chosen such that $F$ is minimized:
\begin{gather}
  \sum_{x'} 2 V(x,x') \Delta(x') = z \delta(x)
  \\
  \Delta(x)
  =
  \frac{z}{2} V^{-1}(x,0)
  \,,
  \quad
  z = \frac{2 \Delta(0)}{V^{-1}(0,0)}
  \,,
\end{gather}
where the inverse $V^{-1}$ is the quasi-1D Ginzburg-Landau propagator.

For a long SNS junction, the supercurrent is proportional to the overlap of
the tails of $\Delta$ from the S-leads:
\begin{align}
  \Delta(x)
  &\simeq
  e^{-i\varphi/2}
  \frac{\Delta_LV^{-1}(x,0)}{V^{-1}(0,0)}
  +
  e^{i\varphi/2}
  \frac{\Delta_RV^{-1}(x,L)}{V^{-1}(0,0)}
  \,,
  \\
  \label{eq:F-res}
  F
  &\simeq
  \mathrm{const.}
  +
  \cos(\varphi) \frac{\hbar}{2e} I_{c,\rm int}
  \,,
  \\
  I_{c,\rm int}
  &\simeq
  \frac{4e}{\hbar}
  \frac{\Delta_L\Delta_R}{|V^{-1}(0,0)|^2} V^{-1}(0,L)
  \,.
\end{align}
In the gradient expansion, we have
\begin{align}
   V(x,x') &= [-\mathcal{A}\frac{D}{2} \partial_x^2 + \mathcal{A} b] \delta(x-x')
   \,,
   \\
   V^{-1}(x,x') &= \frac{L_g}{\mathcal{A} D} e^{-|x-x'|/L_g}
   \,,
   \\
   I_c
   &=
   8 \mathcal{A} b \Delta_L \Delta_R L_g e^{-L/L_g}
   \,,
\end{align}
and $L_g=\sqrt{D/(2b)}$.

\section{Interactions in a Creutz-ladder SNS junction}

As an example, we show how the interaction-mediated current appears in the Creutz-ladder.
The interacting 1D Creutz-ladder flat-band model is
\begin{align}
  H_0(k)
  &=
  2t
  \begin{pmatrix}
    \cos (ka)  &  -i\sin(ka)
    \\
    i\sin(ka) & -\cos (ka)
  \end{pmatrix}
  \\
  &=
  2t V_k \sigma_z V_k^\dagger
  \,,
  \\
  V_k
  &=
  \frac{1}{2}
  \begin{pmatrix}
    e^{-ika} + 1 & e^{-ika} - 1
    \\
    e^{-ika} - 1 & e^{-ika} + 1
  \end{pmatrix}
  \\
  &=
  \begin{pmatrix}
    u_k & v_k
  \end{pmatrix}
  \,,
  \\
  H_{\rm BdG}(n,n')
  &=
  \begin{pmatrix}
    H_0(n,n') - \mu & \Delta_n \delta_{nn'}
    \\
    \Delta_n^* \delta_{nn'} & -H_0(n,n') + \mu
  \end{pmatrix}
  \,. \label{equation:CreutzBdG}
\end{align}
Here, $n$ is the unit cell index, $\sigma_z=\diag(1,-1)$ in the band indices, and $a$ is the distance between unit cells. We have assumed that the positions of the orbitals A and B are at the same location, which makes the quantum metric of this system equal to the minimal quantum metric~\cite{huhtinen2022,Tam2023}. With such a choice, any spatially dependent phase factors of the order parameter must be the same for the two orbitals; this justifies the assumption in Eq.~\eqref{equation:CreutzBdG} that the order parameters of the two orbitals have the same amplitude and phase.
Fourier transforms are as $f_k=\sum_n e^{ikan}f_n$, $f_n=\sum_k e^{-ikan}f_k$,
$\sum_k\equiv\int_{-\pi/a}^{\pi/a}\frac{a\dd{k}}{2\pi}$.

The Ginzburg--Landau expansion of the mean-field free energy then is,
\begin{align}
  F
  &=
  \sum_{nn'} V(n,n')\Delta(n)\Delta(n')
  \,,
  \\
  V(n,n')
  &=
  \frac{2}{U}\delta_{nn'}
  -
  \Pi(n,n')
  \,.
\end{align}
Here $U$ is the interaction energy.

The electronic part is
\begin{align}
  \Pi(n,n')
  &=
  T
  \sum_\omega \tr\{ G_0(n'-n) \bar{G}_0(n-n')\}
  \,,
\end{align}
where $G_0 = [i\omega - H_0+\mu]^{-1}$, $\bar{G}_0 = [i\omega + H_0 - \mu]^{-1}$.
This gives
\begin{align}
  \Pi(n,n')
  &=
  \sum_{kk'} e^{-i(k-k')(n-n')a}
  \sum_{\alpha\alpha'=\pm}
  \\\notag
  &\times
  \tr V_k\frac{1 + \alpha\sigma_z}{2} V_{k}^\dagger V_{k'}\frac{1 + \alpha'\sigma_z}{2} V_{k'}^\dagger
  \\\notag
  &\times
  T\sum_\omega \frac{1}{(i\omega - 2 t \alpha + \mu)(i\omega + 2 t \alpha' - \mu)}
  \,.
\end{align}
Let us now set $\mu=-2t$, and restrict to the lower flat band $\alpha=\alpha'=-1$,
dropping other terms. The neglected parts are small in $T/\mu\ll1$.

Then, taking the Fourier transform
\begin{align}
  \Pi(q)
  &=
  \frac{1}{4T}
  \sum_{k} |v_k^\dagger v_{k-q}|^2
  \,,
\end{align}
where $\sum_\omega \frac{T}{\omega^2}=1/(4T)$.
For small $q$, this is directly related to the (minimal) quantum metric,
\begin{align}
  \Pi(q)
  &\simeq
  \frac{1}{4T}
  -
  \frac{1}{4T}
  \xi_{g}^2
  q^2
  \,,
  &
  \xi_g^2
  &=
  \int_{-\pi/a}^{\pi/a}
  \frac{a \dd{k}}{2\pi} g(k)
  \,.
\end{align}
From the above, this is a general feature of flat-band models~\cite{chen2024}. Note that for our choice of orbital positions, the quantum metric of the system is automatically the minimal quantum metric. For arbitrary orbital positions, one should do this calculation by considering the orbital positions explicitly, to obtain the dependence on the minimal quantum metric as in Ref.~\cite{hu2024anomalouscoherencelengthsuperconductors}.

The Fourier-transformed Ginzburg-Landau propagator is then
\begin{align}
  \label{eq:creutz-glpropagator}
  V^{-1}(q)
  &=
  \frac{
    1
  }{
    \frac{2}{U} - \Pi(q)
  }
  \simeq
  \frac{
    4T
  }{
    \frac{8T}{U} - 1 + \xi_g^2q^2
  }
  \,.
\end{align}
The mean-field superconducting transition is at $T_c=U/8$.
The quadratic expansion in $q$ implies $V^{-1}(x)\propto{}e^{-x/L_g}$,
where $L_g=\xi_g\sqrt{T_c/(T - T_c)}$.

\begin{figure}[t]
  \includegraphics{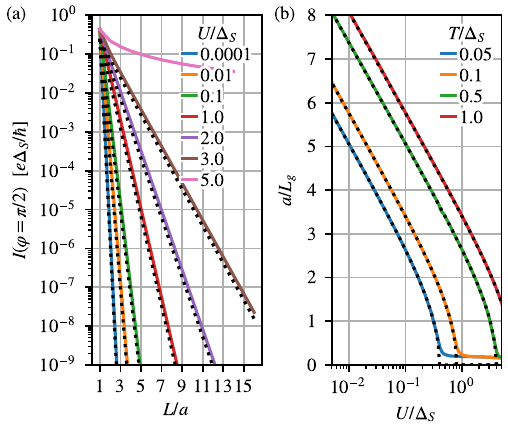}
  \caption{\label{fig:creutz-sc}
    (a)
    Supercurrent in S/Creutz-N/S junction vs.~length, for different interaction
    strengths, at temperature $T=\Delta_S/2$.
    The chemical potential $\mu=-2t$ is at a flat band, and $t=50\Delta_S$.
    The critical interaction strength is $U_c=4\Delta_S$.
    Solid lines: numerical self-consistent BdG calculation.
    Black dotted: Eq.~\eqref{eq:Icreutz}.
    (b)
    Supercurrent decay length $L_g$, $I_s\propto{}e^{-L/L_g}$, vs. interaction strength at different temperatures.
    Solid lines: the fitted slope of the numerically computed $-\ln{}I_s(L)$ at large $n$.
    Black dotted: Eq.~\eqref{eq:Lgcreutz}.
  }
\end{figure}

Since $v_k$ has a very simple form in the Creutz ladder,
we can also calculate the result
without the $q\to0$ expansion:
\begin{align}
  \Pi(q)
  &=
  \sum_k \cos^2(qa/2) = \cos^2(qa/2)
  \,.
\end{align}
Note that this means $\Pi(n)=0$ for $|n|>1$, exhibiting the compact
localization properties of an exact flat band, visible in the behavior of
$G_0$.  This however does not imply that the propagator $V^{-1}$ is
similarly localized.

Writing $z=T/T_c>1$, the propagator is
\begin{align}
  V^{-1}(n)
  &=
  4T
  \int_{-\pi/a}^{\pi/a}\frac{a\dd{q}}{2\pi}
  \frac{
    e^{-iqna}
  }{
    z - \cos^2(qa/2)
  }
  \\
  &=
  \frac{4Te^{-|n| a / L_g}}{\sqrt{z(z-1)}}
  \,,
  \;
  L_g
  =
  \frac{a}{\log[2z - 1 + 2\sqrt{z(z-1)}]}
  \,,
  \label{eq:Lgcreutz}
\end{align}
as found by changing variables $q=ia\ln z$ and calculating the residue.

The interaction-mediated supercurrent in the Creutz ladder is then
\begin{align}
  \label{eq:Icreutz}
  I_{c,\rm int}
  &=
  \frac{\Delta_L\Delta_R}{2T} \sqrt{z (z - 1)} e^{-L/L_g}
  \,,
\end{align}
within the Gingzburg-Landau expansion and assuming $\mu=-2t$, $|t|\gg{T}$.

Comparison of the above to numerical calculations without the Ginzburg-Landau
approximation are shown in Fig.~\ref{fig:creutz-sc}.  It assumes that
$\Delta_n$ is fixed to $\Delta_S e^{-i\varphi/2}$ for $n\le0$,
$\Delta_S e^{i\varphi/2}$ for $n\ge{}L/a$, and
$\Delta_n=U\langle{c_{n\uparrow}c_{n\downarrow}}\rangle$ for $0<n<L/a$
are determined self-consistently.

\begin{figure}[t]
  \includegraphics{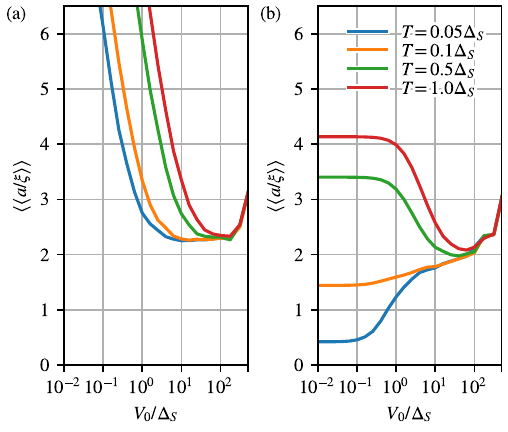}
  \caption{\label{fig:creutz-sc-dis}
    (a)
    Disorder-averaged supercurrent inverse decay length for a noninteracting Creutz ladder, i.e., with $U=0$,
    in the presence of potential disorder of amplitude $V_0$.
    Here $t=t'=50\Delta_S$, but the results do not change
    for larger values.
    (b)
    Same for $U/\Delta_S=0.5$, and corresponding intrinsic
    $T_c=0.0625\Delta_S$.
  }
\end{figure}

\section{Effect of disorder on the coherence length in isolated flat bands}

Disorder changes the precise conditions needed for destructive interference that creates localization in a flat band. It can increase the decay length of the supercurrent
and it modifies the effective superfluid weight \cite{pyykkonen2023-snq,lau2022}.
This requires the disorder to be strong enough,
here of the same order as the superconducting energy scales. 

The effect is
illustrated in Fig.~\ref{fig:creutz-sc-dis} for the Creutz ladder, where numerically computed
disorder average of the inverse decay length $\langle\langle \xi_{\rm decay}^{-1}
\rangle\rangle$ is plotted, in the presence of random uniformly
distributed on-site potential disorder $V_{\rm dis}\in [-V_0, V_0]$.
The decay length $\xi_{\rm decay}$ is defined by fitting the SNS
supercurrent length dependence to $e^{-L/\xi}$ for large $L$, where the behavior follows such an exponential.
Note that this definition excludes the short-distance behavior.

For $V_0\to0$, the inverse decay length $\xi_{\rm decay}^{-1}$ in the
noninteracting case is infinite, due to the compact localization of
the exactly flat bands: as we show in the main text, for exactly flat bands with no dispersive bands nearby and no interactions, the proximity effect (pairing amplitude induced by the superconductor) vanishes exactly after a short distance, instead of an exponential decay. Therefore, in the Creutz ladder, which has two exact flat bands and no dispersive ones, the long-distance decay length $\xi_{\rm decay}$ defined by the exponential decay becomes zero for $V_0=0$ (this analysis does not capture the short-distance coherence length of the order of one unit cell).   For finite $V_0$, the inverse decay length
becomes finite. It decreases until it saturates for $V_0>\Delta_S$ toward a
constant nonzero value of the order of the lattice constant, as
expected for a disordered 1D system. Interactions are required for it
to decrease below this limit. At large $V_0\gtrsim{}t$ where energy differences between sites are large compared to hopping, inverse decay length increases again.

With interactions present [Fig.~\ref{fig:creutz-sc-dis}(b)], increasing disorder strength makes the
decay length to converge toward the noninteracting disordered value in this system. Superfluid weight in s-wave flat-band superconductors is
known to be suppressed by the disorder in the same way as the intrinsic $\Delta$
\cite{lau2022}, and a similar effect is seen in the low-temperature
results of Fig.~\ref{fig:creutz-sc-dis}(b) (orange and blue curves), namely that disorder decreases the coherence length.  At high temperatures (green and red curves), the result at low disorder is essentially the same as for the interacting clean system, while for stronger disorder, it approaches the behavior of the disordered non-interacting system (Fig.~\ref{fig:creutz-sc-dis}(a)). The results show that the effect of temperature is much stronger than disorder in this system, and below the scale $V_0 \sim 0.5 \Delta_S$, disorder effects are negligible.

\subsection{Effect of disorder on the coherence length in the presence of dispersive bands}

\begin{figure}[t]
  \includegraphics{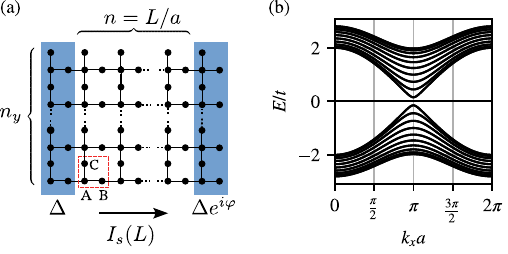}
  \caption{\label{fig:liebstrip-setup}
    (a)
    S/Lieb-N/S junction, consisting
    of a Lieb lattice strip $n_y$ unit cells (dashed square) wide.
    Superconducting pair
    potential is $\Delta\ne0$ at B and C sites of the edges (``S'' part) and $\Delta=0$
    on the $n$ unit cells that make up the ``N'' part.
    Phase difference across the ``S'' edges drives supercurrent $I_s(L)\propto{}e^{-L/\xi}$
    through the system.
    (b)
    Spectrum of a strip $n_y=10$ unit cells wide, as a function of momentum
    $k_x$ along the strip, for $\Delta=0$.
  }
\end{figure}

We also consider the effect of disorder in a case where
the system has some dispersive bands (we focus here on the case of no interactions in the normal (N) part). For this purpose, consider
a quasi-1D system: Lieb lattice strip that is $n_y$ unit cells
wide, see Fig.~\ref{fig:liebstrip-setup}. The intra-unit cell hoppings ($K$)
and the hopping matrices $J_x$ for $(i,j)\mapsto(i+1,j)$, $J_y$ for
$(i,j)\mapsto(i,j+1)$ are
\begin{align}
  K &= t\begin{pmatrix} 0 & 1 & 1 \\ 1 & 0 & 0 \\ 1 & 0 & 0 \end{pmatrix}
  \,,
  \\
  J_x &= t\begin{pmatrix} 0 & 1 & 0 \\ 0 & 0 & 0 \\ 0 & 0 & 0 \end{pmatrix}
  \,,
  &
  J_y &= t\begin{pmatrix} 0 & 0 & 1 \\ 0 & 0 & 0 \\ 0 & 0 & 0 \end{pmatrix}
  \,.
\end{align}
The Bloch Hamiltonian for an infinite lattice is
$H(k_x,k_y)=K+[e^{ik_xa}J_x + e^{ik_ya}J_y + \mathrm{h.c.}]$ and has
eigenvalues $E=\{0, \pm t\sqrt{4+2\cos(k_xa)+2\cos(k_ya)}\}$.  For a
strip with $n_y\gg{}1$ unit cells in $y$-direction, the dispersive
subbands nearest to the flat band $E=0$ have a low-energy expansion
$E^2\simeq{}\frac{\pi^2t^2}{4n_y^2} + t^2a^2(k_x-\pi)^2$ around
$k_x=\pi/a$.  The Dirac cone part of the Lieb lattice spectrum is
gapped due to the finite width of the strip.

As the spectrum aside from the exactly flat band is gapped (assuming
$\mu=0$), the S/Lieb-N/S supercurrent follows a tunneling (WKB-type)
decay at long distances.  This is also seen from Eq.~(8), in the
simplest approximation,
\begin{align}
  f(L)^2
  &\simeq
  \Bigl(
  \int_{-\infty}^{\infty}\frac{dk_x\,e^{-ik_xL}}{\pi^2T^2 + \frac{\pi^2t^2}{4n_y^2} + t^2a^2k_x^2}
  \Bigr)^2
  \propto
  e^{-L/\xi}
  \,.
\end{align}
When the temperature becomes large compared to the gap, $T>\pi t/(2n_y)$,
the result crosses over from the gapped coherence length $\xi_{\rm gap}=n_ya/\pi$
to the Fermi surface coherence length $\xi_{v_F}=v_F/(2\pi T)$
corresponding to the Lieb lattice Dirac cone velocity $v_F=ta$,
\begin{align}
  \label{eq:liebstrip-xian}
  \xi\simeq{}[(\pi/n_y)^2 + (2\pi T/t)^2]^{-1/2} a
\end{align}
This temperature dependence of the inverse decay length is illustrated
in Fig.~\ref{fig:liebstrip-sc-dis}(a). The numerical results are
obtained in a tight-binding model with $\Delta=\Delta_S$ at lattice
sites at the left and right ends of a strip, and $\Delta=0$ on the
sites of the N-part. These results thus again show how in the absence of interactions, flat bands do not contribute to the long-range decay length of the proximity effect.

\begin{figure}[t]
  \includegraphics{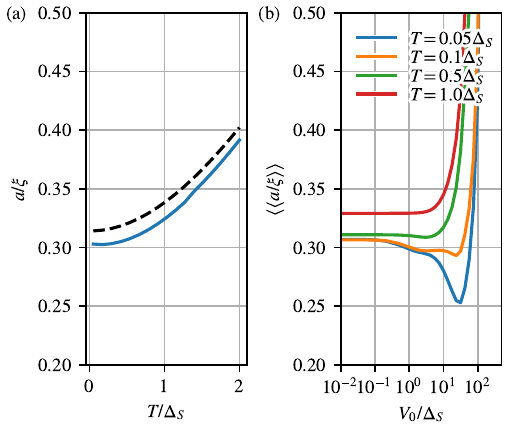}
  \caption{\label{fig:liebstrip-sc-dis}
    (a)
    Supercurrent inverse decay length in S/Lieb-N/S geometry as a function of temperature (at $\varphi=\pi/2$).
    The normal (N) part is a Lieb lattice strip with $n_y=10$ unit cells in the transverse direction.
    Hopping $t=50\Delta_S$ is assumed and $U=0$.
    Solid: numerics. Dashed: Eq.~\eqref{eq:liebstrip-xian}.
   (b)
    Disorder-averaged supercurrent inverse decay length in the same system,
    in the presence of potential disorder of amplitude $V_0$.
  }
\end{figure}

The effect of uniformly distributed potential disorder ($V_{\rm
  dis}\in [-V_0, V_0]$) is illustrated in
Fig.~\ref{fig:liebstrip-sc-dis}(b).  At weak disorder, there is
little change in the inverse decay length, which remains at the value
determined by the energy gap and temperature. At disorder strength
of $V_0\gtrsim{}\Delta_S$ the inverse decay length starts to decrease,
until $V_0\gtrsim{}t$ where it increases again,
qualitatively similarly as for the Creutz ladder in the previous section. 

\section{Computer codes}

Computer codes used in this manuscript and the supplement can be found at
\url{https://gitlab.jyu.fi/jyucmt/2024-flatband-s-junctions}.

\fi

\end{document}